\documentclass[a4paper,10pt]{article}
\usepackage{amsmath,amssymb,amsfonts,amsthm}
\usepackage{tikz}
\usepackage[utf8]{inputenc}
\usepackage[T2A]{fontenc} 
\usepackage{mathtext}
\usepackage[english]{babel}
\usepackage{cmap} 
\usepackage{cite}

\usepackage{csquotes}
\usepackage{ gensymb }
\usepackage[unicode]{hyperref}
\usepackage{ textcomp }
\usepackage{indentfirst}
\usepackage[version=3]{mhchem}
\usepackage{amsmath}
\usepackage{slashed}
\usepackage{float}
\usepackage{bbold}
\usepackage{braket}
\usepackage{ytableau}	

\usepackage{setspace,amsmath}
\usepackage[left=15mm, top=20mm, right=15mm, bottom=20mm, nohead]{geometry} 
\usepackage{graphicx}
\usepackage[nottoc]{tocbibind}
\usepackage{float}
\usepackage{adjustbox}
\DeclareMathOperator{\Tr}{Tr}

\usepackage{graphicx}

\title{\vspace{-1cm}{\Large {\bf 
			Genus expansion of matrix models and $\hbar$ expansion of KP hierarchy
		}
		\date{}
		\author{
			{\bf A. Andreev$^{a,c}$}\thanks{andreev.av@phystech.edu},
			{\bf A. Popolitov$^{a,b,c}$}\thanks{popolit@gmail.com},
			{\bf A. Sleptsov$^{a,b,c}$}\thanks{sleptsov@itep.ru}, 
			{\bf A. Zhabin$^{a,c}$}\thanks{alexander.zhabin@yandex.ru}}
	}}

\begin{document}

\maketitle
\vspace{-5.2cm}

\begin{center}
	\hfill ITEP/TH-15/20 \\
	\hfill IITP/TH-11/20 \\
	\hfill MIPT/TH-10/20 
\end{center}

\vspace{2.7cm}

\begin{center}
	$^a$ {\small {\it Institute for Theoretical and Experimental Physics, Moscow 117218, Russia}}\\
	$^b$ {\small {\it Institute for Information Transmission Problems, Moscow 127994, Russia}}\\
	$^c$ {\small {\it Moscow Institute of Physics and Technology, Dolgoprudny 141701, Russia }}
\end{center}

\vspace{0.5cm}


\begin{abstract}
	We study $\hbar$ expansion of the KP hierarchy following Takasaki-Takebe \cite{TAKASAKI_1995} considering several examples of matrix model $\tau$-functions with natural genus expansion. Among the examples there are solutions of KP equations of special interest, such as generating function for simple Hurwitz numbers, Hermitian matrix model, Kontsevich model and Brezin-Gross-Witten model. We show that all these models with parameter $\hbar$ are $\tau$-functions of the $\hbar$-KP hierarchy and the expansion in $\hbar$ for the  $\hbar$-KP coincides with the genus expansion for these models. Furthermore, we show a connection of recent papers considering the $\hbar$-formulation of the KP hierarchy \cite{Natanzon_2016, kazarian2015combinatorial} with original Takasaki-Takebe approach. We find that in this approach the recovery of enumerative geometric meaning of $\tau$-functions is straightforward and algorithmic.
\end{abstract}

\vspace{.5cm}

	\section{Introduction}	
	Matrix models have a long history of development for over 40 years and by now have found applications in numerous branches of theoretical physics. Among the topics, that have recently attracted much attention, are Jackiw-Teitelboim (JT) gravity, SYK-like models and $ (q,t)$-deformation. JT gravity, which is a 2D dilaton theory of gravity, was found to be dual to Hermitean matrix model in a particular regime \cite{saad2019jt, witten2020volumes, witten2020matrix}. Partition functions of JT gravity on surfaces with certain types of boundary can be calculated explicitly using intersection theory and therefore correspond to the genus expansion of a matrix model. Since the correlators one needs to consider on matrix model side are not usual products of traces, the spectral curve changes its shape. Intense studies of JT gravity are caused by its holographic duality to 1D Schwarzian theory \cite{Jensen_2016, Maldacena_2016_1, Engels_y_2016} which is the low energy limit of SYK model \cite{Kitaev_2018, Maldacena_2016}. SYK, in turn, has its own independent relation to matrix and tensor models \cite{Witten_2019, Gurau_2017}. Tensor models often have the same properties as matrix models, e.g. Feynman diagrams expansion and integrability \cite{amburg2020correspondence,itoyama2020complete,klebanov2017tasi}.   Finally, $ (q,t)$-deformed matrix models have found their use in localization computations in supersymmetric gauge theories. In particular, they are indispensable for calculating supersymmetric Wilson loop averages and proving recursive identities for them \cite{Lodin_2019, Cassia_2019}.
	
	In a certain range of parameters matrix model has a perturbative expansion depicted by Feynman diagrams. Each diagram is a ribbon graph that can be drawn on a certain Riemann surface. It naturally determines a genus decomposition (which corresponds to $ \frac{1}{N} $ decomposition after suitable rescaling of matrix model parameters) for partition function and correlators. One way to approach this genus expansion is via the, so-called, loop equations, which are the consequence of matrix model Ward identities and can be solved recursively. Each higher genus part can be calculated explicitly from genus zero one- and two-point correlators. Under certain mild assumptions this initial data can be repackaged into an algebrogeometric structure on a certain Riemann surface, which is called a spectral curve. Then in terms of this spectral curve the recursion procedure, described by loop equations, acquires universal form, the so-called \textit{spectral curve topological recursion} \cite{chekhov2006free, chekhov2006hermitian, eynard2007invariants, eynard2005topological, alexandrov2004partition, alexandrov2007m, alexandrov2007instantons}. This procedure, thanks to its universality, became very popular in mathematical physics during last years \cite{kramer2019topological,dunin2019loop,dunin2019combinatorial,borot2017blobbed,bychkov2020combinatorics}.

	The notion of the spectral curve first appeared in integrable systems, therefore, matrix models are closely related to various integrable structures, in particular, to the KP/Toda integrable hierarchy. The simplest description of this integrable hierarchy is an infinite set of non-linear differential equations with the first equation given by
	\begin{equation}
		\frac{1}{4} \frac{\partial^{2} F}{\partial t_{2}^{2}} = \frac{1}{3} \frac{\partial^{2} F}{\partial t_{1} \partial t_{3}} - \frac{1}{2} \left( \frac{\partial^{2} F}{\partial t_{1}^{2}} \right)^{2} - \frac{1}{12} \frac{\partial^{4} F}{\partial t_{1}^{4}}
	\end{equation}
	where F-function depends on an infinite set of time variables $ \textbf{t} = \{{t_{1}, t_{2}, t_{3}, \dots}\} $. Partition functions and correlators in various quantum systems are $\tau$-functions of some integrable hierarchy ($ \tau = \exp(F) $) \cite{its1990differential,izergin1992determinant,gorsky1995integrability,gorsky1998multiscale}, which often implies a matrix model description for them. Historically, KP equations appeared in description of nonlinear wave motion in  two-dimensional media. Presently, it is known that the KP hierarchy is connected with many fundamental structures of modern theoretical physics such as infinite-dimensional Lie algebras \cite{date1982transformation, jimbo1983solitons}, projective manifolds \cite{sato1983soliton, segal1985loop}, matrix models of 2D gravity \cite{brezin1993exactly, douglas1990strings, gross1993nonperturbative}, lattice gauge theories of QCD \cite{KHARCHEV_1995, mironov1996unitary}, knot theory \cite{mironov2013genus}, enumerative geometry \cite{Okounkov_2000}, combinatorics \cite{kazarian2015combinatorial} and others.
	
	In papers \cite{takasaki1992quasi, TAKASAKI_1995} it was suggested to introduce a "Planck's constant" $\hbar$ into the KP hierarchy and study the dispersionless limit of the hierarchy at $\hbar \to 0$. This limit can be understood as a quasi-classical limit in 2D gravity, i.e., it is consistent with the genus expansion coming from Virasoro constraints in two-dimensional quantum gravity. By taking $\hbar=1$ one obtains the classical KP hierarchy. After introduction of $\hbar$ KP equations are deformed, for instance, the first equation takes the form
	\begin{equation}
		\frac{1}{4} \frac{\partial^{2} F^{\hbar}}{\partial t_{2}^{2}} = \frac{1}{3} \frac{\partial^{2} F^{\hbar}}{\partial t_{1} \partial t_{3}} - \frac{1}{2} \left( \frac{\partial^{2} F^{\hbar}}{\partial t_{1}^{2}} \right)^{2} - \frac{\hbar^{2}}{12} \frac{\partial^{4} F^{\hbar}}{\partial t_{1}^{4}}.
	\end{equation}
	In the $\hbar \to 0 $ limit it turns into celebrated dispersionless equation \cite{krichever1992dispersionless, dubrovin1992hamiltonian}, provided that the limit for the F-function exists. This explicit introduction of $\hbar$ via rescaling of times is called the $\hbar$-formulation (or expansion) of the KP hierarchy or shortly $\hbar$-KP. We also sometimes call it $\hbar$ deformation of KP hierarchy, though it is not a deformation in common sense. The $\hbar$-formulation can be explicitly performed for all structures in the KP theory: Lax operator, $ W $-symmetries and an element of the $ GL(\infty) $ group. Following Takasaki-Takebe, Natanzon and Zabrodin recently introduced formulation of $\hbar$-KP \cite{Natanzon_2016} in terms of common equations on deformed $\tau$-functions and F-functions ($ F^{\hbar} = \hbar^{2} \log \tau^{\hbar} $, note extra factor $\hbar^{2}$ for correct limit $\hbar \to 0$) with one more parameter $ x $, which is the shift of first variable $ t_{1} \rightarrow t_{1} + x $. They managed to obtain explicit solution for the F-function in terms of Cauchy-like data and explicit combinatorial constants.

	
	Formal $\hbar$-KP suffers from the lack of explicit examples, so our first goal is to introduce a set of solutions of the $\hbar$-KP which are obtained by inserting $\hbar$-dependence into some very well known solutions of the usual KP. The choice of these examples is not accidental. There are some solutions of KP equations of special practical interest, such as Kontsevich model \cite{kontsevich1992intersection}, Brezin-Gross-Witten partition function \cite{gross1980possible, brezin1980external}, generating function for simple Hurwitz numbers \cite{alexandrov2012integrability} and Hermitian matrix model \cite{gerasimov1991matrix, kharchev1991matrix, kharchev1993generalized}. All these models play an important role both in modern high energy physics and contemporary mathematics. All considered models have a matrix model representation and, as we discussed above, there is the natural genus expansion (consistent with the topology of the corresponding invariants) coming from the combinatorics of associated Feynman graphs for each model. Therefore, our second goal is to investigate, how the expansion in $\hbar$ for $\hbar$-KP matches with the genus expansion. We explicitly demonstrate that they do coincide. We do this explicitly in great detail, so as to dispel any doubts. Finally, we show that recent papers considering $\hbar$-formulation of KP hierarchy \cite{Natanzon_2016, kazarian2015combinatorial} do coincide with original Takasaki-Takebe deformation.
	
	Moreover, in considered examples with explicit free-fermion formalism we observe that
	the insertion of $\hbar$ is \textit{algorithmic}. One just needs to put $\hbar$ in front of differentiation (that is, "momentum operator") and $1/\hbar$ in front of the exponent (that is "the action"). This simple rescaling, motivated by quantum mechanics in a very straightforward way,
	reproduces the correct genus expansion. Thus, one of the main points of this paper is that if one has a KP $\tau$-function for which one does not know the enumerative geometric meaning of its different components, and even how to split the function into the components of different genera, then a way to reveal it may be through this above mentioned procedure. 
	
	Let us now explain, why the examples considered are important. We start with the celebrated Kontsevich model. This matrix model is a generating function for intersection numbers of $\psi$-classes:
	\begin{equation}
		Z_{K}(\Lambda) = \frac{\int DX \exp\left( \Tr\left( i\frac{X^{3}}{3!} + \frac{\Lambda X^{2}}{2} \right) \right)}{\int DX \exp \left( \Tr \frac{\Lambda X^{2}}{2} \right)} \sim \exp \left( \left\langle \exp\left( \sum_{m=0}^{\infty} T_{m} \psi_{m} \right) \right\rangle \right)
	\end{equation}
	From this point of view it is a partition function of 2D topological gravity \cite{kontsevich1992intersection}. According to Witten's conjecture \cite{witten1990two} it coincides with a partition function of physical 2D quantum gravity. The key to the proof of this conjecture is that Kontsevich model is a $\tau$-function of the KdV hierarchy \cite{Kharchev_1992, kontsevich1992intersection}, which can be obtained as a reduction of the KP hierarchy. In this work we are interested in the genus expansion of Kontsevich model, which allows us to separate contributions of Riemann surfaces of different genera.
	
	Another example of great interest in physics is Brezin-Gross-Witten model. Starting point of this model is a partition function of lattice QCD with Wilson action studied by Brezin and Gross \cite{brezin1980external}, Gross and Witten \cite{gross1980possible}:
	\begin{equation}
		Z_{BGW}(J,J^{+}) = \int DU \exp(\Tr(J^{\dagger}U + JU^{\dagger}))\:,
	\end{equation}
	where the integration is over unitary matrices. Complex matrix $ J $ is understood as an "external field". The model depends only on the eigenvalues of the matrix $ JJ^{\dagger} $. Choice of variables $ t_{k} = \Tr(JJ^{\dagger})^{k} $ or $ t_{-2k+1} = -\frac{1}{2k-1} \Tr(JJ^{\dagger})^{-k+1/2} $ corresponds to weak and strong-field limits, which are called "character" phase and "Kontsevich" phase respectively. BGW model is a particular example of Generalized Kontsevich Model and its partition function is a $\tau$-function of KP hierarchy \cite{mironov1996unitary}. Virasoro constraints provide genus decomposition for multiresolvents and F-function \cite{Alexandrov_2009}.
	
	Hurwitz numbers are counting ramified coverings of Riemann sphere, and were originally studied by Hurwitz in 19-th century. These unsophisticated quantities appeared to have deep connections to both mathematical and physical structures. On the one hand, the generating function for simple Hurwitz numbers is the $\tau$-function of the KP hierarchy. On the other hand, it was shown in \cite{rusakov1990loop} that 2D Yang-Mills partition function with the gauge group $U(N)$ is precisely the generating function for simple Hurwitz numbers. This correspondence is very fruitful and has been widely discussed in literature \cite{gross1993two,kostov1997two,kimura2008holomorphic,griguolo2005double}. In particular, from the physics point of view to obtain a contribution of a particular genus one just considers Yang-Mills theory on a surface of this genus. Various physical phenomena, for example, Douglas-Kazakov phase transition, have natural explanation in terms of the genus expansion of Hurwitz numbers \cite{douglas1993large}. 
	
	In knot theory the Ooguri-Vafa (OV) partition function for HOMFLY polynomials of some knot is the Hurwitz partition function \cite{mironov2013genus,mironov2013genus2,sleptsov2014hidden}. Thus Hurwitz numbers have, even if implicit and peculiar, connection with 3D Chern-Simons theory. The OV partition function for any torus knot is the $\tau$-function of the KP hierarchy \cite{mironov2013character}, while for an arbitrary knot only the large $N$ limit of the partition function is the KP $\tau$-function and already first few corrections violate the integrability  \cite{mironov2013genus}. Moreover, Hurwitz numbers are of great interest for Gromov-Witten theory, which was pointed out by Okounkov and Pandharipande \cite{okounkov2006gromov}, and makes yet another connection with physics and string theory. While in the broad sense Hurwitz numbers are integrable, that is, there is an explicit Frobenius formula that expresses any given Hurwitz number through group theory quantities, in a more narrow sense only certain subfamilies of Hurwitz numbers are integrable, that is, are $\tau$-functions of KP hierarchy. It is known, for instance, that Hurwitz partition function with at least 3 generic ramification profiles is not a KP $\tau$-function, and it is an open question to develop a (generalized) theory of integrability, which would naturally contain all Hurwitz partition functions. In this paper we are concentrated on the generating function for simple Hurwitz numbers. Genus expansion of the generating function separates contributions of coverings of fixed genera.
	
	Matrix models were known to be KP solutions since development of the 2D gravity. Another example we consider is Hermitian matrix model:
	\begin{equation}\label{hmm}
		Z_{N}(\textbf{t}) = \frac{\int DX \exp \left(-\frac{1}{2} \Tr(X^{2}) + \sum_{k=1}^{\infty} t_{k} \Tr(X^{k})\right)}{\int DX \exp \left(-\frac{1}{2} \Tr(X^{2}) \right)}.
	\end{equation}
	Perturbative solution for this model is well-known \cite{Mironov_2017} and takes the form of \textit{dessin d'enfant}: ribbon graphs on Riemann surfaces. Insertion of the parameter $\hbar$ into \eqref{hmm} allows us to distinguish surfaces of different genera. Note that Hermitian matrix model is integrable for an arbitrary, not necessary Gaussian, action. 
	
	The paper is organized as follows. In section 2 we introduce the notation of all well-known objects which we are going to use in our calculations, such as Schur polynomials, $ \widehat{gl(\infty)} $ algebra and free fermions. Section 3 is devoted to classical KP hierarchy and to the important class of solutions which arises in examples -- hypergeometric $\tau$-functions. In section 4 we consider already mentioned $\hbar$-formulation of KP hierarchy. The main propositions that will be in use are presented. In section 5 we explicitly introduce solutions of $\hbar$-KP. For each model we explain insertion of parameter and prove that they are, indeed, solutions of $\hbar$-KP. Generalization of deformation to the entire hypergeometric family of solutions is discussed in section 6. Finally, in section 7 structure of general $\hbar$-KP solutions with parameters $\hbar$ and $x$ are studied.
	
	\section{Boson-fermion correspondence}
	In this section we review the notions of Young diagrams, Schur polynomials, Fock space and free fermions, which form essential vocabulary for the KP theory.
	
	\subsection{Schur polynomials}
	Extensive information about Schur polynomials can be found in \cite{macdonald1998symmetric}, here we summarize only what we do need. Let us consider an ordered set of non-negative integers $ \lambda_{1} \ge \lambda_{2} \ge \dots \ge \lambda_{l} \ge 0 $. We will denote this set by $ \lambda = [\lambda_{1}, \lambda_{2}, \dots, \lambda_{l}] $ and call it a Young diagram. Each Young diagram corresponds to a partition of an integer $ |\lambda| := \lambda_{1} + \lambda_{2} + \dots + \lambda_{l} $ onto $ l(\lambda) $ non-zero parts $ \lambda_{i} $. Graphical representation of Young diagrams is a finite collection of boxes, arranged in left-justified rows, with length of each row equal to $ \lambda_{1}, \lambda_{2}, \dots, \lambda_{l} $. For example, the diagram $ [5,3,2] $:
	\begin{equation}
		\begin{ytableau}
			\; & \; & \; & \; & \;\cr
			\; & \; & \;\cr
			\; & \;\cr
		\end{ytableau}
	\end{equation}
	
	Let $ \textbf{t} = \{t_{1}, t_{2}, \dots\} $ be an infinite set of variables. Schur polynomials $ s_{\lambda}(\textbf{t}) $ are enumerated by Young diagrams and are defined via determinant formula:
	\begin{equation}\label{Schur_polynomials_definition}
		s_{\lambda}(\textbf{t}) = \det_{1 \le i,j \le l(\lambda)} h_{\lambda_{i} - i + j}(\textbf{t}),
	\end{equation}
	where polynomials $ h_{k}(\textbf{t}) $ are defined with the help of generating function:
	\begin{equation}
		\exp \left( \sum_{k=1}^{\infty}t_{k}z^{k} \right) = \sum_{k=0}^{\infty} h_{k}(\textbf{t}) z^{k}
	\end{equation}
	Let $ h_{k}(\textbf{t}) = 0 $ for each $ k < 0 $. It is clear from the definition \eqref{Schur_polynomials_definition} that if $ l(\lambda) = 1 $, then $ s_{[k]}(\textbf{t}) = h_{k}(\textbf{t}) $. We will call $ s_{[k]}(\textbf{t}) $ \textit{symmetric} Schur polynomials.
	
	First few examples of Schur polynomials are:
	\begin{equation}
		\begin{gathered}
			s_{\varnothing}(\textbf{t}) = 1,\\
			s_{[1]}(\textbf{t}) = t_{1},\\
			s_{[2]}(\textbf{t}) = \frac{t_{1}^{2}}{2} + t_{2}, \;\;\; s_{[1,1]}(\textbf{t}) = \frac{t_{1}^{2}}{2} - t_{2}\\
			s_{[3]}(\textbf{t}) = \frac{t_{1}^{3}}{6} + t_{1}t_{2} + t_{3}, \;\;\; s_{[2,1]}(\textbf{t}) = \frac{t_{1}^{3}}{3} - t_{3}, \;\;\; s_{[1,1,1]}(\textbf{t}) = \frac{t_{1}^{3}}{6} - t_{1}t_{2} + t_{3}
		\end{gathered}
	\end{equation}
	
	One more fact we will need is the Cauchy-Littlewood completeness identity
	\begin{equation}\label{Cauchy-Littlewood_identity}
		\sum_{\lambda} s_{\lambda}(\textbf{t}) s_{\lambda}(\textbf{t}^{'}) = \exp \left( \sum_{k=1}^{\infty} k t_{k} t_{k}^{'} \right)
	\end{equation}
	that is, among other things, useful in proving character expansion formulas.
	
	\subsection{Fock space, free fermions and $ \widehat{gl(\infty)} $ algebra}
	Again, here we summarize only what we find necessary, more details can be found in \cite{kac2013bombay}. There is a natural way to describe solutions of KP hierarchy in terms of free fermions. First of all, let us introduce an infinite dimensional Clifford algebra with generators $\{ \psi_{n}, \psi_{m}^{*}| n,m \in \mathbb{Z} \}$ and commutation relations:
	\begin{equation}\label{Clifford_algebra}
		\{ \psi_{n}, \psi_{m} \} = 0, \;\;\; \{ \psi_{n}^{*}, \psi_{m}^{*} \} = 0, \;\;\; \{ \psi_{n}, \psi_{m}^{*} \} = \delta_{n,m}
	\end{equation}
	Introduce generating series for fermions:
	\begin{equation}\label{fermionic_free_fields}
		\psi(z) = \sum_{k \in \mathbb{Z}} \psi_{k} z^{k}, \;\;\;\;\;\; \psi^{*}(z) = \sum_{k \in \mathbb{Z}} \psi_{k}^{*} z^{-k-1}
	\end{equation}
	
	Fermionic Fock space is defined by the action of Clifford algebra on vacuum vector $\ket{0}$, which can be understood as "Dirac sea". Action of fermionic generators on vacuum vector is
	\begin{equation}
		\psi_{k} \ket{0} = 0, \;\;\; k < 0, \;\;\;\;\;\;\;\;\;\; \psi_{k}^{*} \ket{0} = 0, \;\;\; k \ge 0 
	\end{equation}
	With respect to vacuum $ \ket{0} $, operators $ \psi_{k}, \; k \ge 0 $ and $ \psi_{k}^{*}, \; k < 0 $ are creation operators, while $ \psi_{k}, \; k < 0 $ and $ \psi_{k}^{*}, \; k \ge 0 $ are annihilation operators. Dual vacuum vector $ \bra{0} $ (the covacuum) has the properties
	\begin{equation}
		\bra{0} \psi_{k}^{*} = 0, \;\;\; k < 0, \;\;\;\;\;\;\;\;\;\;  \bra{0} \psi_{k} = 0, \;\;\; k \ge 0 
	\end{equation}
	
	We denote a normal ordering of fermionic operators as $ \colon (\dots) \colon $. All annihilation operators are moved to the right and all creation operators to the left, with respect to $ (-1) $ with each transposition of fermions. For example, $ \colon \psi_{1}^{*} \psi_{1} \colon = -\psi_{1} \psi_{1}^{*} $, $ \colon \psi_{-1} \psi_{0} \colon = - \psi_{0} \psi_{-1} $. Note that it is not the same as transposition of fermions with the help of commutation relations \eqref{Clifford_algebra} ($ \colon \psi_{1}^{*} \psi_{1} \colon \ne \colon (1 - \psi_{1} \psi_{1}^{*}) \colon $)
	
	Bosonic Fock space $ \mathcal{B}^{(0)} $ is a linear space of polynomials of an infinite number of time variables $ t_{1}, t_{2}, t_{3}, \dots $ ($ \mathcal{B}^{(0)} := \mathbb{C}[t_{1}, t_{2}, t_{3}, \dots] $). Schur polynomials are basis vectors in this linear space. For the following explanation we will need a space 
	\begin{equation}
		\mathcal{B} = \bigoplus_{k \in \mathbb{Z}} z^{k} \mathcal{B}^{(k)} = \bigoplus_{k \in \mathbb{Z}} z^{k} \mathbb{C}[t_{1}, t_{2}, t_{3}, \dots]
	\end{equation}

	Linear spaces $ \mathcal{B} $ и $ \mathcal{F} $ can be endowed with rich algebraic structure. Namely, consider a Lie algebra of matrices $ gl(\infty) $: each matrix $ A \in gl(\infty) $ is an infinite matrix, with additional requirement that only finitely many diagonals are non-zero, $ A_{ij} = 0 $ for $ |i - j| \gg 0 $ is satisfied. A standard basis for the algebra consists of matrices $ E_{ij} $, that have 1 on $ i,j $'s place and 0 everywhere else. It is easy to show that $ E_{ij} $ satisfy standard commutation relations
	\begin{equation}
		[E_{ij}, E_{kl}] = \delta_{jk} E_{il} - \delta_{il} E_{kj}
	\end{equation}
	We will need $ gl(\infty) $'s central extension: the $ \widehat{gl(\infty)} $ in order to define an action on the fermionic Fock space. As a linear space, central extension $ \widehat{gl(\infty)} $ is $ gl(\infty) \oplus \mathbb{C}c $. Commutator of two arbitrary elements $ A, B \in \widehat{gl(\infty)} $ is given by
	\begin{equation}\label{gl_infinity_commutation_relations}
		[A, B] = AB - BA + \alpha(A,B)c
	\end{equation}
	where $ \alpha(A,B) $ is linear in each variable and therefore can be defined on the basis elements $ E_{ij} $:
	\begin{equation}
		\begin{cases}
		\alpha(E_{ij}, E_{ji}) = -\alpha(E_{ji}, E_{ij}) = \mathbb{1}, & \text{if} \;\; i \le 0, j \ge 1;\\
		\alpha(E_{ij}, E_{mn}) = 0, & \text{otherwise},
		\end{cases}
 	\end{equation}
	where $\mathbb{1}$ is the identity element. Now the $ \widehat{gl(\infty)} $ representation on space $ \mathcal{F} $ is, in terms of basis elements:
	\begin{equation}
		r(E_{ij}) = \colon \psi_{i} \psi_{j}^{*} \colon
	\end{equation}
	It is straightforward to check that $ r $ is indeed a representation: commutation relations \eqref{gl_infinity_commutation_relations} for $ r(A), r(B) $ are preserved. The central charge $ c $ in this representation is equal to 1. 
	
	Now let us consider matrices
	\begin{equation}
		H_{n} = \sum_{k \in \mathbb{Z}} \colon \psi_{k} \psi_{k+n}^{*} \colon
	\end{equation}
	which satisfy the following commutation relations
	\begin{equation}
		[H_{n}, H_{m}] = n\delta_{n,-m}
	\end{equation}
	and thus generate the Heisenberg subalgebra $ \mathcal{A} \subset \widehat{gl(\infty)} $. 
	
	It is possible to construct a map $ \Phi: \mathcal{F} \rightarrow \mathcal{B} $, which is a homomorphism of representations. Under this homomorphism the operators $ H_{k} $ map into multiplication and differentiation w.r.t. times
	\begin{equation}
		\begin{cases}
		H_{k} \rightarrow \frac{\partial}{\partial t_{k}} \\
		H_{-k} \rightarrow kt_{k}\\
		H_{0} \rightarrow \mu \mathbb{1}
		\end{cases}
	\end{equation}
	The map $\Phi$ is called the boson-fermion correspondence. In what follows we will consider KP hierarchy, thus we are interested only in results of the action of $ GL(\infty) $ on vacuum vector $ \ket{0} $. Group $ GL(\infty) $ is a standard exponential map from algebra $ \widehat{gl(\infty)} $. Map $ \Phi $ in this case can be explicitly written as vacuum expectation value:
	\begin{equation}
		\Phi(G\ket{0}) = \bra{0} e^{H(\textbf{t})} G \ket{0}
	\end{equation}
	where $ H(\textbf{t}) $ is a sum of elements of commutative subalgebra of $ \mathcal{A} $
	\begin{equation}
		H(\textbf{t}) = \sum_{k=1}^{\infty} t_{k} H_{k}
	\end{equation}
	
	\section{KP hierarchy}
	In this section we briefly review the main facts about KP equations and solutions. For the detailed explanation see \cite{miwa2000solitons}. KP hierarchy is an infinite set of non-linear differential equations with the first equation given by
	\begin{equation}
		\frac{1}{4} \frac{\partial^{2} F}{\partial t_{2}^{2}} = \frac{1}{3} \frac{\partial^{2} F}{\partial t_{1} \partial t_{3}} - \frac{1}{2} \left( \frac{\partial^{2} F}{\partial t_{1}^{2}} \right)^{2} - \frac{1}{12} \frac{\partial^{4} F}{\partial t_{1}^{4}}
	\end{equation}
	It is more common to work with $ \tau $-function $ \tau(\textbf{t}) = \exp(F(\textbf{t})) $ than with free energy $ F(\textbf{t}) $. We assume that $ \tau(\textbf{t}) $ is at least a formal power series in times $ t_{k} $, and maybe it is even a convergent series. Entire set of equations of hierarchy can be written in terms of $ \tau $-function using Hirota bilinear identity, which, in turn, is equivalent to the following functional equation
	\begin{equation}\label{Hirota_functional_relation}
		(z_{1} - z_{2})\tau^{[z_{1},z_{2}]}\tau^{[z_{3}]} + (z_{2} - z_{3})\tau^{[z_{2},z_{3}]}\tau^{[z_{1}]} + (z_{3} - z_{1})\tau^{[z_{3},z_{1}]}\tau^{[z_{2}]} = 0
	\end{equation}
	where
	\begin{equation}\label{time_shift_classical}
		\begin{gathered}
			\tau^{[z_{1}, \dots, z_{m}]}(\textbf{t}) = \tau \left( \textbf{t} + \sum_{i=1}^{m}[z_{i}^{-1}] \right)\\
			\textbf{t} + [z^{-1}] = \left\{ t_{1} + \frac{1}{z}, t_{2} + \frac{1}{2z^{2}}, t_{3} + \frac{1}{3z^{3}}, \dots \right\}
		\end{gathered}
	\end{equation}
	Equation \eqref{Hirota_functional_relation} should be satisfied for arbitrary $ z_{1}, z_{2}, z_{3} $. One can expand $ \tau $-function at the vicinity of $ z_{i} = \infty $ and obtain partial differential equation for $ \tau $-function at every term $ z_{1}^{-k_{1}} z_{2}^{-k_{2}} z_{3}^{-k_{3}} $.
	
	On the one hand, all formal power series solutions of KP hierarchy can be decomposed over the basis of Schur polynomials
	\begin{equation}\label{bosonic_representation}
		\tau(\textbf{t}) = \sum_{\lambda} C_{\lambda} s_{\lambda}(\textbf{t}).
	\end{equation}
	Function written as a formal sum over Schur polynomials is a KP solution if and only if coefficients $ C_{\lambda} $ satisfy the Pl\"{u}cker relations. The first such relation is
	\begin{equation}\label{plucker_relations}
		C_{[2,2]} C_{[\varnothing]} - C_{[2,1]} C_{[1]} + C_{[2]} C_{[1,1]} = 0.
	\end{equation}
	We call $ \tau $-function, written in the form \eqref{bosonic_representation}, the $ \tau $-function in \textit{bosonic represenation}. 
	
	On the other hand, $ \tau $-function is an image under boson-fermion correspondence of a point on the orbit of the vacuum $ \ket{0} $ under the action of some element $ G $ of $ GL(\infty) $:
	\begin{equation}\label{fermionic_representation}
		\tau(\textbf{t}) = \bra{0} e^{H(\textbf{t})} G \ket{0}
	\end{equation}
	We call $ \tau $-function, written in the form \eqref{fermionic_representation}, the $ \tau $-function in \textit{fermionic representation}. Comparing bosonic \eqref{bosonic_representation} and fermionic \eqref{fermionic_representation} representations of a $\tau$-function we read
	\begin{equation}\label{connection_between_bosonic_and_fermionic_representations}
		C_{\lambda} = \det_{l(\lambda) \times l(\lambda)} \left( G_{i_{-1}, i_{-2}, i_{-3}, \dots}^{-1, -2, -3, \dots} \right)
	\end{equation}
	In the last formula one has to calculate determinant of the matrix obtained by intersection of columns $ -1, -2, -3, \dots $ and rows $ i_{-1}, i_{-2}, i_{-3}, \dots $ of the initial matrix $ G $. Integers $ i_{-k} $ are determined from Young diagram $ \lambda = [i_{-1} + 1, i_{-2} + 2, \dots] $. Thus, each concrete $ C_{\lambda} $ is a determinant of a finite-dimensional $ l(\lambda) \times l(\lambda) $ matrix.
	
	\subsection{Hypergeometric $\tau$-functions}
	In this paper we are interested in a subset of KP $\tau$-functions: the $\tau$-functions of hypergeometric type, or simply hypergeometric $\tau$-functions. This relatively simple set of KP solutions contains surprisingly rich amount of physical examples. It was first introduced in \cite{KHARCHEV_1995} and further developed in \cite{orlov2001hypergeometric}. In fermionic representation these $\tau$-functions have the form:
	\begin{equation}\label{hypergeometric_tau_function_fermionic_representation}
		\begin{gathered}
			\tau(\textbf{t}) = \bra{0} e^{H(\textbf{t})} e^{A(\beta)} \ket{0}\\
			A(\beta) = \sum_{k=1}^{\infty} \beta_{k} A_{k}, \;\;\;\;\;\; A_{k} = \sum_{n \in \mathbb{Z}} r(n)r(n-1) \dots r(n-k+1) \colon \psi_{n} \psi_{n-k}^{*} \colon
		\end{gathered}
	\end{equation}
	where $ r(n) $ is an arbitrary function and $ \beta = \{ \beta_{1}, \beta_{2}, \dots \} $ is an arbitrary set of parameters. Matrix $ A_{k} $ has non-zero elements of a specific form on the $ (-k)$-th diagonal. It is easy to obtain another form of the matrix $ A_{k} $ which is more convenient in some cases:
	\begin{equation}\label{A_k_matrix}
		A_{k} = \oint \frac{dz}{2 \pi i} \colon \left[ \left(\frac{1}{z} r(D)\right)^{k} \psi(z) \right] \cdot \psi^{*}(z) \colon
	\end{equation}
	where $ D = z \frac{d}{dz} $, therefore, $ r(D)z^{n} = r(n)z^{n} $. Using the explicit form of the fermionic fields \eqref{fermionic_free_fields} one can obtain \eqref{A_k_matrix}.
	
	Bosonic representation of hypergeometric $\tau$-functions requires a notion of a content $ c(w) $ of a box $ w $ of Young diagram $ \lambda $:
	\begin{equation}
		c(w) = j - i, \;\; 1 \le i \le l(\lambda), \;\; 1 \le j \le \lambda_{i}.
	\end{equation}
	For example, boxes of the diagram $ [5,3,2] $ have following contents:
	\begin{equation}
		\begin{ytableau}
			0 & 1 & 2 & 3 & 4\cr
			-1 & 0 & 1\cr
			-2 & -1\cr
		\end{ytableau}
	\end{equation}
	Hypergeometric $ \tau $-functions in bosonic representation are given by
	\begin{equation}\label{hypergeometric_tau_function_bosonic_representation}
		\tau(\textbf{t}) = \sum_{\lambda} r_{\lambda} s_{\lambda}(\beta) s_{\lambda}(\textbf{t}),
	\end{equation}
	where $ s_{\lambda}(\beta) $ is a Schur polynomial of variables $ \beta_{k} $ and
	\begin{equation}
		r_{\lambda} = \left( \prod_{w \in \lambda} r(c(w)) \right).
	\end{equation}

	\section{$\hbar$-formulation of the KP hierarchy}
	In this section we review papers, where $\hbar$-formulation of the hierarchy was considered.
	
	\subsection{Takasaki-Takebe formulation}
	Takasaki and Takebe have introduced $\hbar$-formulation of the  KP hierarchy in a series of papers \cite{takasaki1992quasi,TAKASAKI_1995} by insertion of a formal parameter $ \hbar $. The main idea was to study the dispersionless KP hierarchy at the limit $ \hbar \rightarrow 0 $. Existence of the limit imposes restrictions on free energy $ F $: parameter $ \hbar $ should be inserted into the logarithm of $ \tau $-function and equations "properly" -- there must not be any negative powers of $ \hbar $. Here we use slightly different notation from \cite{TAKASAKI_1995}, namely, connection between $F$-function and $\tau$-function is given by
	\begin{equation}
		F^{\hbar}(\textbf{t}) = \hbar^{2} \log\left(\tau^{\hbar}(\textbf{t})\right).
	\end{equation}
	The first equation of the dispersionless hierarchy takes the form:
	\begin{equation}\label{dispersionless_equation}
		\frac{1}{4} \frac{\partial^{2} F}{\partial t_{2}^{2}} = \frac{1}{3} \frac{\partial^{2} F}{\partial t_{1} \partial t_{3}} - \frac{1}{2} \left( \frac{\partial^{2} F}{\partial t_{1}^{2}} \right)^{2}
	\end{equation}
	
	The proper insertion of parameter $\hbar$ into $\tau$-functions is given by the following proposition.
	
	\textbf{Proposition 1. \cite[Proposition 1.7.15]{TAKASAKI_1995}} \textit{$ \tau $-functions with "good" quasi-classical behaviour have the form
		\begin{equation}\label{Takasaki_Takebe_deformation}
			\begin{gathered}
				\tau^{\hbar}(\textbf{t}) = \bra{0} e^{H(\textbf{t}/\hbar)} \exp\left( \frac{1}{\hbar} A^{\hbar} \right) \ket{0}\\
				A^{\hbar} = \oint \frac{dz}{2 \pi i} \colon \left[\hat{A} \left(z,\hbar \frac{d}{dz}\right) \psi(z) \right] \cdot \psi^{*}(z) \colon
			\end{gathered}
		\end{equation}
		where $ \hbar $ appears in the differential operator $ \hat{A}\left(z, \hbar \frac{d}{dz}\right) $ in front of the differentiation and also may occur in the expansion coefficients
		\begin{equation}
			\hat{A}\left(z, \hbar \frac{d}{dz}\right) = \sum_{i \in \mathbb{Z}, j \ge 0} a_{i,j}(\hbar) z^{i} (\hbar \partial_{z})^{j}
		\end{equation}
		Each $ a_{i,j}(\hbar) $ contains only non-negative powers of $\hbar$
		\begin{equation}\label{expansion_coefficients}
			a_{i,j}(\hbar) = \sum_{m=0}^{\infty} a_{i,j}^{(m)} \hbar^{m}
		\end{equation}}
	In all examples we consider in this paper, coefficients $ a_{i,j} $ turn out to be independent of $ \hbar $. Arbitrary $\tau$-function of the deformed hierarchy we examine in the next subsection \eqref{bosonic_representation_deformed}. Obviously, putting $\hbar \to 1$ restores the original formulation of the KP hierarchy.
	
	
	
	
	
	\subsection{Natanzon-Zabrodin formulation}
	Natanzon and Zabrodin were following Takasaki-Takebe to introduce their rescaling of the KP hierarchy \cite{Natanzon_2016}. This expansion allows $ F $-functions to depend on arbitrary powers of formal parameter $ \hbar $. Deformed hierarchy is defined via functional equation \eqref{Hirota_functional_relation} but the shift of variables of $\tau$-function \eqref{time_shift_classical} changes to explicitly include $\hbar$:
	\begin{equation}\label{time_shift_deformed}
		\begin{gathered}
			(z_{1} - z_{2})\tau^{[z_{1},z_{2}]}\tau^{[z_{3}]} + (z_{2} - z_{3})\tau^{[z_{2},z_{3}]}\tau^{[z_{1}]} + (z_{3} - z_{1})\tau^{[z_{3},z_{1}]}\tau^{[z_{2}]} = 0\\
			\tau^{[z_{1}, \dots, z_{m}]}(\textbf{t}) = \tau \left( \textbf{t} + \hbar \sum_{i=1}^{m}[z_{i}^{-1}] \right)\\
			\textbf{t} + \hbar [z^{-1}] = \left\{ t_{1} + \frac{\hbar}{z}, t_{2} + \frac{\hbar}{2z^{2}}, t_{3} + \frac{\hbar}{3z^{3}}, \dots \right\}
		\end{gathered}
	\end{equation}
	The first equation of the hierarchy is then of the form
	\begin{equation}\label{first_KP_equation_deformed}
		\frac{1}{4} \frac{\partial^{2} F}{\partial t_{2}^{2}} = \frac{1}{3} \frac{\partial^{2} F}{\partial t_{1} \partial t_{3}} - \frac{1}{2} \left( \frac{\partial^{2} F}{\partial t_{1}^{2}} \right)^{2} - \frac{\hbar^{2}}{12} \frac{\partial^{4} F}{\partial t_{1}^{4}}
	\end{equation}
	Note that at the limit $ \hbar \rightarrow 0 $ this equation is exactly dispersionless equation \eqref{dispersionless_equation}.
	
	
	We call the deformed hierarchy $ \hbar $-KP. Note that by rescaling of times $ t_{k} \rightarrow \frac{t_{k}}{\hbar} $ in any $\tau$-function of a classical KP hierarchy \eqref{Hirota_functional_relation} one obtains $\tau$-function of $\hbar$-formulation of KP. So, at first glance, the change from KP is trivial. Unfortunately, such a trivial rescaling of times does not yield any good properties to $ F $-function. Non-trivial deformation occurs when parameter $ \hbar $ appears in Pl\"{u}cker coefficients $ C_{\lambda} $ in \eqref{bosonic_representation}. In fermionic representation it is stated how $\hbar$ should be inserted into an element of the group $ GL(\infty) $ \eqref{Takasaki_Takebe_deformation}, therefore, via the formula \eqref{connection_between_bosonic_and_fermionic_representations} it is possible to restore $ C_{\lambda}^{\hbar} $. In this case functions $ F^{\hbar}(\textbf{t}) $ may have correct quasi-classical limit. Moreover, these functions may have geometrical structure such as genus expansion (we will see it in the next section on various examples). However, this deformation does not say how to deform an arbitrary $\tau$-function to obtain good quasi-classical behaviour. Summarizing, any $\tau$-function of $\hbar$-KP is given by
	\begin{equation}\label{bosonic_representation_deformed}
		\tau^{\hbar}(\textbf{t}) = \sum_{\lambda} C_{\lambda}^{\hbar} s_{\lambda}\left(\frac{\textbf{t}}{\hbar}\right).
	\end{equation}
	
	A statement that it is possible to obtain deformed $\tau$-function from the classical one with the change of variables $ t_{k} \rightarrow \frac{t_{k}}{\hbar} $ is correct vice versa. Deformed $ \tau $-function $ \tau^{\hbar}(\textbf{t}) $ will be a solution of classical KP after the inverse change of variables, therefore, coefficients $ C_{\lambda}^{\hbar} $ should satisfy the classical Pl\"{u}cker relations. The mentioned ideas can be formulated in the following proposition.
	
	\textbf{Proposition 2. ($ \hbar $-KP solution criterion)} \textit{$ \tau$-function of the form \eqref{bosonic_representation_deformed} satisfies $ \hbar $-KP equations if and only if coefficients $ C_{\lambda}^{\hbar} $ satisfy the classical Pl\"{u}cker relations.}
	
	\vspace{0.5cm}
	Main purpose of the $\hbar$-formulation was to obtain explicit formula for the $ F $-function which is absent in classical KP. It is necessary to add one more parameter for this reason. It is known \cite{miwa2000solitons} that in KP theory variable $ t_{1} $ is distinguished -- $ \tau $-function will remain a solution of hierarchy after shift of first variable $ t_{1} \rightarrow t_{1} + x $. $ \tau $-functions are then $ \tau^{\hbar}(x, \textbf{t}) = f(x) \hat{\tau}(x + t_{1}, t_{2}, t_{3}, \dots) $, where $f(x)$ is an arbitrary smooth function that does not violate KP equations. All solutions of the hierarchy can be written in the following way.
	
	\textbf{Proposition 3. \cite[Theorem 2.1]{Natanzon_2016}} \textit{Let $ \tau^{\hbar}(x,\textbf{t}) = f(x)\hat{\tau}(x+t_{1}, t_{2}, \dots) $ be a $ \tau $-function of $ \hbar $-KP and function $ \tau^{\hbar}(x, \textbf{0}) $ is infinitely differentiable with respect to variable $ x $. The coefficients of the series
	\begin{equation}\label{bosonic_representation_with_parameter_x}
		\tau^{\hbar}(x,\textbf{t}) = \sum_{\lambda} C_{\lambda}^{\hbar}(x) s_{\lambda}\left(\frac{\textbf{t}}{\hbar}\right)
	\end{equation}
	are connected by determinant relations
	\begin{equation}\label{coefficient_determinant_relations}
		C_{\lambda}^{\hbar}(x) = (C_{0}^{\hbar}(x))^{1 - l(\lambda)} \underset{1 \le i, j \le l(\lambda)}{\det} \left[ \sum_{k=0}^{j - 1} (-\hbar)^{k} \binom{j -1}{k} \partial_{x}^{k} C_{\lambda_{i} - i + j - k}^{\hbar}(x) \right]
	\end{equation}
	where $ \binom{j-1}{k} = \frac{(j-1)!}{k!(j-1-k)!} $ are binomial coefficients, $ C_{0}^{\hbar}(x) = C_{\varnothing}^{\hbar}(x), C_{k}^{\hbar}(x) = C_{[k]}^{\hbar}(x) $}
	
	\textit{Conversely, if $ C_{k}^{\hbar}(x) $ are arbitrary infinitely differentiable functions with respect to variable $ x $ and $ C_{\lambda}^{\hbar} $ are defined via relations \eqref{coefficient_determinant_relations} then series of the form \eqref{bosonic_representation_with_parameter_x} is a solution of $ \hbar $-KP ($\hbar \ne 0$).}
	
	Therefore $ \tau $-function in this form is determined by an arbitrary set of infinitely differentiable functions $ C_{k}^{\hbar}(x) $. To avoid confusion with Pl\"{u}cker coefficients the explicit dependence on the variable $ x $ is emphasized. Similarly to ordinary differential equations the set of functions $ C_{k}^{\hbar}(x) $ is called Cauchy-like data (initial data at $ \textbf{t} = 0 $). They are connected with $\tau$-function by the following relations
	\begin{equation}
		\tau^{\hbar}(x,\textbf{0}) = C_{0}^{\hbar}(x), \;\;\; \partial_{k}^{\hbar} \tau^{\hbar}(x,\textbf{t}) \Big|_{\textbf{t} = 0} = \frac{k}{\hbar} C_{k}^{\hbar}(x)
	\end{equation}
	where the deformed partial derivative $ \partial_{k}^{\hbar} $ is defined via symmetric Schur polynomials:
	\begin{equation}
		\partial_{k}^{\hbar} = \frac{k}{\hbar} s_{[k]}(\hbar \widetilde{\partial}), \;\;\;\;\; \widetilde{\partial} = \left\{ \partial_{1}, \frac{1}{2} \partial_{2}, \frac{1}{3} \partial_{3}, \dots \right\}
	\end{equation}
	First few deformed derivatives: $ \partial_{1}^{\hbar} = \partial_{1}, \partial_{2}^{\hbar} = \partial_{2} + \hbar \partial_{1}^{2}, \partial_{3}^{\hbar} = \partial_{3} + \frac{3}{2} \hbar \partial_{1} \partial_{2} + \frac{1}{2} \hbar^{2} \partial_{1}^{3} $. At $ \hbar = 0 $ deformed derivatives transform into usual ones $ \partial_{k}^{\hbar = 0} = \partial_{k} $.
	
	The next step is to obtain explicit formula for the function $ F^{\hbar} = \hbar^{2} \log \tau^{\hbar} $. For this reason one will need deformed variables $ t_{\lambda}^{\hbar} $ and combinatorial coefficients $ P_{\lambda}^{\hbar} $. Detailed description of these objects can be found in \cite{Natanzon_2016}.
	
	\textbf{Proposition 4. \cite[Theorem 4.3]{Natanzon_2016}} \textit{For arbitrary set of smooth functions}
	\textit{$$ \textbf{f} = \{f_{0}^{\hbar}(x), f_{1}^{\hbar}(x), \dots\} $$
		there exists a unique solution $ F^{\hbar}(x,\textbf{t}) $ of the $\hbar$-KP hierarchy such that $ F^{\hbar}(x,0) = f_{0}(x) $ and $ \partial_{k}^{\hbar} F^{\hbar}(x,\textbf{t})\lvert_{\textbf{t}=0} = f_{k}^{\hbar}(x) $. This solution is of the form
		\begin{equation}\label{F_function_solution}
			F^{\hbar}(x,\textbf{t}) = f_{0}^{\hbar}(x) + \sum_{|\lambda| \ge 1} \frac{f_{\lambda}^{\hbar}(x)}{\sigma(\lambda)} t_{\lambda}^{\hbar}
		\end{equation}
		where $ f_{[k]}^{\hbar}(x) = f_{k}^{\hbar}(x) $ and 
		\begin{equation}\label{f_coefficients}
			f_{\lambda}^{\hbar}(x) = \sum_{m \ge 1} \sum_{\substack{s_{1} + l_{1} + \dots + s_{m} + l_{m} = |\lambda|\\ 1 \le s_{i}; 1 \le l_{i} \le l(\lambda) - 1}} P^{\hbar}_{\lambda}
			\begin{pmatrix}
				s_{1} \dots s_{m} \\
				l_{1} \dots l_{m}
			\end{pmatrix}
			\partial_{x}^{l_{1}} f_{s_{1}}^{\hbar}(x) \dots \partial_{x}^{l_{m}} f_{s_{m}}^{\hbar}(x)
		\end{equation}
		for $ l(\lambda) > 1 $. $ \sigma(\lambda) = \prod_{i\ge1}m_{i}! $, where exactly $ m_{i} $ parts of the partition $\lambda$ have length $ i $.}
	
	The solution in terms of logarithm of the $\tau$-function is determined by arbitrary set of functions $ \textbf{f} $ as well. These functions are called Cauchy-like data for the $ F $-function. Formula \eqref{F_function_solution} is nothing else but modified Taylor series in variables $ t_{k} $ with parameters $ x $ and $ \hbar $. The main result of this proposition is the existence of universal combinatorial coefficients $ P_{\lambda}^{\hbar} $ which depend only on the form of KP equations and are independent of particular solution. These coefficients allow us to construct entire solution from Cauchy-like data.
		

	\section{Examples of $\hbar$-KP solutions}

	One of the drawbacks of Takasaki-Takebe original development is the lack of examples. In this section we consider some examples, chosen in such a way that they have a well-known geometrical interpretation of genus expansion. We show that all of these functions are solutions of $\hbar$-KP explicitly. 
	For some examples we consider both fermionic and bosonic representation to compare with $\hbar$-formulation in terms \eqref{Takasaki_Takebe_deformation} and \eqref{bosonic_representation_deformed}.

	\subsection{Hurwitz numbers $\tau$-function}
	Genus expansion of the generating function for simple Hurwitz numbers was studied in \cite{bouchard2007hurwitz, mironov2009virasoro}. Insertion of rescaling parameter $\hbar$ into the function leads to the deformation of KP equations. In this subsection we are following review \cite{kazarian2015combinatorial}. Hurwitz numbers as a solution of $\hbar$-KP is presented in section 5.1.3. To understand geometrical deformation we briefly recall the definition of Hurwitz numbers. 

	\subsubsection{Hurwitz numbers and classical Hurwitz $\tau$-function}
	Hurwitz numbers were originally studied by Hurwitz in the 19-th century. Simple Hurwitz numbers are counting ramified coverings of Riemann sphere by two-dimensional surface of genus $ g $ with $ m $ simple ramification points and one point with ramification profile given by partition $ \mu = [\mu_{1}, \mu_{2}, \dots, \mu_{l(\mu)}] $. It turns out that this count can be expressed purely in terms of symmetric group theory (that is, the only non-trivial data are the monodromies around ramification points). We will denote simple Hurwitz numbers as $ h_{m;\mu}^{\circ} $. Then the following expression holds \cite{kazarian2015combinatorial}:
	\begin{equation}\label{simple_Hurwitz_numbers}
		h_{m;\mu}^{\circ} = \frac{1}{|\mu|!} \big| \{(\eta_{1},\dots,\eta_{m}), \eta_{i} \in C_{2}(S_{|\mu|}): \eta_{m} \circ \dots \circ \eta_{1} \in C_{\mu}(S_{|\mu|})\} \big|
	\end{equation}
	where $ S_{|\mu|} $ is the symmetric group of permutations of $ |\mu| $ elements, $ C_{2}(S_{|\mu|}) $ is a set of all transpositions in $ S_{|\mu|} $ and $ C_{\mu}(S_{|\mu|}) $ is a set of all permutations of a cycle type $ \mu $. $ \eta_{1}, \dots, \eta_{m} $ correspond to simple ramification points and their product corresponds to the distinguished ramification point with $ \mu $-monodromy.
	
	Connected simple Hurwitz numbers $ h_{m;\mu} $ are defined similarly but covering surface must be connected. From the combinatorial point of view definition \eqref{simple_Hurwitz_numbers} stays almost the same but now it is necessary to count only such transpositions $ \eta_{i} $, whose generated subgroup \textlangle $ \eta_{1}, \dots, \eta_{m} $\textrangle $ \subset S_{|\mu|} $ is transitive.
	
	Generating function for simple Hurwitz numbers is a classical KP $\tau$-function (see for example \cite{Okounkov_2000})
	\begin{equation}
		\tau_{H}(\textbf{t}) = \sum_{m = 0}^{\infty} \sum_{\mu} h_{m;\mu}^{\circ} t_{\mu_{1}} t_{\mu_{2}} \dots t_{\mu_{l(\mu)}} \frac{u^{m}}{m!},
	\end{equation}
	which we call just "Hurwitz numbers" for simplicity.
	
	General relation between connected and disconnected objects is (connected) = $ \log $(disconnected). Logarithm of $\tau$-function is then a generating function for connected Hurwitz numbers and a solution of KP equations:
	\begin{equation}
		F_{H}(\textbf{t}) = \log(\tau_{H}(\textbf{t})) = \sum_{m = 0}^{\infty} \sum_{\mu} h_{m;\mu} t_{\mu_{1}} t_{\mu_{2}} \dots t_{\mu_{l(\mu)}} \frac{u^{m}}{m!}
	\end{equation}
	
	The generating function for simple Hurwitz numbers belongs to the set of hypergeometric $ \tau $-functions and can be written in the form \eqref{hypergeometric_tau_function_bosonic_representation}
	\begin{equation}\label{tau_hurwitz_classical}
		\tau_{H}(\textbf{t}) = \sum_{\lambda} e^{u c(\lambda)} s_{\lambda}(\beta_{k} = \delta_{k,1}) s_{\lambda}(\textbf{t})
	\end{equation}
	where $ c(\lambda) = \sum_{w \in \lambda} c(w) $. Parameters of the generating function in the set of hypergeometric functions are
	\begin{equation}\label{hurwitz_parameters_as_hypergemetric_function}
		\begin{gathered}
			r(n) = e^{un},\\
			\beta_{1} = 1,\\
			\beta_{k} = 0, k \ge 2
		\end{gathered}
	\end{equation}
	
	\subsubsection{Genus decomposition and deformation of the KP hierarchy}
	As it was mentioned, Hurwitz numbers are counting ramified coverings of Riemann sphere. Genus of the covering surface is defined by ramification parameters via celebrated Riemann-Hurwitz formula. It is better to talk about Euler characteristic of connected coverings which are given by the series $ F_{H}(\textbf{t}) $. In the case of ramified covering with $ m $ simple points of ramification and one point with ramification given by partition $ \mu $ Riemann-Hurwitz formula has the form
	\begin{equation}\label{Riemann_Hurwitz_formula}
		2g - 2 = m - |\mu| - l(\mu)
	\end{equation}
	Now it is possible to distinguish contributions of each genus $ g $ in the generating function. Each point of simple ramification adds +1 to the power of parameter $ \hbar $, each cycle of length $ \mu_{i} $ lowers power of the parameter on $ (\mu_{i} + 1) $. Then we obtain the following change of variables:
	\begin{equation}\label{hurwitz_change_of_variables}
		\begin{gathered}
			t_{\mu_{i}} \rightarrow \hbar^{-\mu_{i} - 1} t_{\mu_{i}}\\
			u \rightarrow \hbar u
		\end{gathered}
	\end{equation}
	Let us multiply the function by $ \hbar^{2} $ to remove negative powers of $ \hbar $ originating from the extra "2" in \eqref{Riemann_Hurwitz_formula}. Then the function $ F_{H}^{\hbar}(\textbf{t}) $ acquires the genus decomposition:
	\begin{equation}
		F_{H}^{\hbar}(\textbf{t}) = \sum_{g = 0}^{\infty} \hbar^{2g} F_{H}^{g}(\textbf{t})
	\end{equation}
	
	Deformed function will not satisfy classical KP equations anymore. One has to perform rescaling of times $ t_{k} $ in KP equations as in \eqref{hurwitz_change_of_variables}. Then the first equation of the deformed hierarchy will be exactly equation \eqref{first_KP_equation_deformed} of formal $\hbar$-KP. Limit $ \hbar \rightarrow 0 $ leads to dispersionless equation \eqref{dispersionless_equation}. We can now guess that parameters $\hbar$ in formal $\hbar$-KP and "topological" approach coincide. However to claim that deformed Hurwitz numbers is a solution of $\hbar$-KP one has to check all the infinite number of KP equations. The proof will be presented in the next section.
	
	Kazarian and Lando perform the same deformation for some other $\tau$-functions of hypergeometric type with an arbitrary $ r(n) $ but with fixed parameters $ \beta_{1} = 1, \beta_{k} = 0, k \ge 2 $ which they call Orlov-Scherbin family of solutions. Let $ r(n) $ be an arbitrary series
	\begin{equation}
		r(n) = a_{0} + a_{1}n + a_{2}n^{2} + \dots
	\end{equation}
	According to Kazarian-Lando, all $\tau$-functions of Orlov-Scherbin family should be deformed in the following way
	\begin{equation}\label{Kazaryan_Lando_deformation}
		\begin{gathered}
			r(c(w)) \rightarrow r(\hbar c(w))\\
			t_{k} \rightarrow \frac{t_{k}}{\hbar^{k+1}},
		\end{gathered}
	\end{equation}
	so that the free energy has a decomposition over even non-negative powers of $ \hbar $ only. We consider the entire hypergeometric family of solutions in section 6.

	\subsubsection{Hurwitz numbers as a solution of $\hbar$-KP}
	Let us show that topological deformation \eqref{hurwitz_change_of_variables} of Hurwitz numbers is a solution of $\hbar$-KP and it coincides with Takasaki-Takebe deformation. Using \eqref{hurwitz_change_of_variables} and explicit bosonic representation \eqref{tau_hurwitz_classical} one gets:
	\begin{equation}
		\tau_{H}^{\hbar}(\textbf{t}) = \sum_{\lambda} e^{u\hbar c(\lambda)} s_{\lambda}(\beta_{k} = \delta_{k,1}) s_{\lambda}\left( \frac{t_{1}}{\hbar^{2}}, \frac{t_{2}}{\hbar^{3}}, \frac{t_{3}}{\hbar^{4}}, \dots \right)
	\end{equation}
	Note that for every term $ \prod_{i=1}^{k} t_{m_{i}} $ in Schur polynomial we have $ \sum_{i=1}^{k} m_{i} = |\lambda| $. Hence
	\begin{equation}
		\prod_{i=1}^{k} \frac{t_{m_{i}}}{\hbar^{m_{i} + 1}} = \frac{1}{\hbar^{|\lambda|}} \prod_{i=1}^{k} \frac{t_{m_{i}}}{\hbar}.
	\end{equation}
	and deformed $\tau$-function can be written as \eqref{bosonic_representation_deformed}:
	\begin{equation}
		\tau_{H}^{\hbar}(\textbf{t}) = \sum_{\lambda} \frac{e^{u\hbar c(\lambda)} s_{\lambda}(\beta_{k} = \delta_{k,1})}{\hbar^{|\lambda|}} s_{\lambda}\left( \frac{\textbf{t}}{\hbar} \right)
	\end{equation}

	According to proposition 2, it is enough to show that coefficients $ C_{\lambda}^{\hbar} $ satisfy the Pl\"{u}cker relations. Non-deformed coefficients  $ C_{\lambda} $ have the form:
	\begin{equation}\label{plucker_coefficients_hurwitz_classical}
		C_{\lambda} = e^{uc(\lambda)} s_{\lambda}(\beta_{k}=\delta_{k,1})
	\end{equation}
	and satisfy the Pl\"{u}cker relations for all $u$. Hence, rescaling of $u$ by $\hbar$ does not change the Pl\"{u}cker relations. Rescaling of $ C_{\lambda} $ by $ \frac{1}{\hbar^{|\lambda|}} $ does not change them either since relations are homogeneous by the sum $ |\lambda_{1}| + |\lambda_{2}| = const $. As a result $ C_{\lambda}^{\hbar} $ satisfy the classical Pl\"{u}cker relations and, hence, deformed $\tau$-function of Hurwitz numbers is a solution of $\hbar$-KP.

	Note here, that deformed Hurwitz $\tau$-function can be rewritten in a more convenient way:
	\begin{equation}\label{tau_hurwitz_deformed}
		\tau_{H}^{\hbar}(\textbf{t}) = \sum_{\lambda} e^{u\hbar c(\lambda)} s_{\lambda}\left(\frac{1}{\hbar}, 0, 0, \dots \right) s_{\lambda}\left( \frac{\textbf{t}}{\hbar} \right)
	\end{equation}
	with rescaled coefficients:
	\begin{equation}\label{plucker_coefficients_hurwitz_deformed}
		C_{\lambda}^{\hbar} = e^{u \hbar c(\lambda)} s_{\lambda}\left( \beta_{k}=\frac{\delta_{k,1}}{\hbar} \right)
	\end{equation}
	This is equivalent to deformation:
	\begin{equation}\label{hurwitz_correct_deformation}
		\boxed{
			\begin{gathered}
			r(n) = e^{un} \rightarrow r(\hbar n) = e^{u \hbar n}\\
			\beta_{1} \rightarrow \frac{\beta_{1}}{\hbar}\\
			t_{k} \rightarrow \frac{t_{k}}{\hbar}
			\end{gathered}}
	\end{equation}

	Now let us consider the deformation in fermionic representation. Since we know parameteres $\mathbf{\beta}$ of $\tau$-function in the Orlov-Scherbin family of solutions, we obtain fermionic representation immediately using \eqref{hypergeometric_tau_function_fermionic_representation} and \eqref{A_k_matrix}:
	\begin{equation}\label{A_hurwitz_classical}
		A_{H} = \sum_{k \in \mathbb{Z}} \beta_{k} A_{k} = \sum_{n \in \mathbb{Z}} e^{un} \colon \psi_{n} \psi_{n-1}^{*} \colon = \oint \frac{dz}{2 \pi i} \colon \left[ \left(\frac{1}{z} \exp(D)\right) \psi(z) \right] \cdot \psi^{*}(z) \colon
	\end{equation}
	where $ D = z \frac{d}{dz} $. From deformation of the form \eqref{hurwitz_correct_deformation}, it is clear how to deform the matrix in fermionic representation: rescaling $ \beta_{1} \rightarrow \frac{\beta_{1}}{\hbar} $ corresponds to the factor $ \frac{1}{\hbar} $ before the integral and rescaling $ e^{un} \rightarrow e^{u \hbar n} $ corresponds to the derivative rescaling $ D = z \frac{d}{dz} \rightarrow \hbar D = z \hbar \frac{d}{dz} $. One obtains a deformed matrix
	\begin{equation}
		A_{H}^{\hbar} = \frac{1}{\hbar} \sum_{n \in \mathbb{Z}} e^{u \hbar n} \colon \psi_{n} \psi_{n-1}^{*} \colon = \frac{1}{\hbar} \oint \frac{dz}{2 \pi i} \colon \left[ \left(\frac{1}{z} \exp(\hbar D)\right) \psi(z) \right] \cdot \psi^{*}(z) \colon
	\end{equation}

	As a result, taking into account $ t_{k} \rightarrow \frac{t_{k}}{\hbar} $, we obtain fermionic representation of $\tau$-function:
	\begin{equation}
		\tau_{H}^{\hbar}(\textbf{t}) = \bra{0} e^{H(\textbf{t}/\hbar)} e^{A_{H}^{\hbar}} \ket{0}
	\end{equation}
	Thus, the topological deformation is consistent with the Takasaki-Takebe deformation, and the expansion coefficients $ a_{i,j}(\hbar) $ \eqref{expansion_coefficients} actually do not depend on $\hbar$.

	\subsection{Hermitian matrix model}
	Let us consider another example of a $ \tau$-function, which is a solution of $ \hbar $ -KP with a "good" quasi-classical behaviour: the partition function for Hermitian Gaussian matrix model. For the Hermitian matrix model, there is a geometrical deformation -- genus expansion: Feynman diagrams in the model are ribbon graphs on a two-dimensional surfaces. It means we can distinguish the contributions of surfaces of different genera. Hermitian matrix model as a solution of $\hbar$-KP is presented in section 5.2.2.
	
	\subsubsection{Classical Hermitian matrix model as $\tau$-function}
	Partition function of Hermitian matrix model (HMM) is
	\begin{equation}\label{hmm_classical}
		Z_{N}(\textbf{t}) = \frac{\int DX \exp \left(-\frac{1}{2} \Tr(X^{2}) + \sum_{k=1}^{\infty} t_{k} \Tr(X^{k})\right)}{\int DX \exp \left(-\frac{1}{2} \Tr(X^{2}) \right)},
	\end{equation}
	where we integrate over all Hermitian matrices $X$ of size $ N\times N $. It is known that HMM partition function can be rewritten as \cite{Mironov_2017}:
	\begin{equation}\label{hmm_schur_expansion}
		Z_{N}(\textbf{t}) = \sum_{\lambda} \frac{s_{\lambda}(\beta_{n} = \frac{1}{2}\delta_{n,2}) }{s_{\lambda}(t_{n}^{''} = \delta_{n,1})} s_{\lambda}(t_{n}^{'} = N/n) s_{\lambda}(\textbf{t}).
	\end{equation}
	This property is called character expansion and has deep impact on integrability \cite{morozov2018q,mironov2017complete,mironov2017correlators,mironov2018sum,cordova2018orbifolds}.

	There is a well-known relation between dimension $ D_{\lambda}(N) $ of representation $ \lambda $ of the group $ GL(N) $ and $ d_{\lambda} $: dimension of representation $ \lambda $ of the symmetric group $ S_{|\lambda|} $ multiplied by $ (1/|\lambda|!) $ \cite{fulton2013representation}. Both these quantities are expressed via Schur polynomials: 
	\begin{equation}\label{dimensions_ratio}
		\frac{D_{\lambda}(N)}{d_{\lambda}} \equiv \frac{s_{\lambda}(t_{n}^{'} = N/n)}{s_{\lambda}(t_{n}^{''} = \delta_{n,1})} = \prod_{w \in \lambda} (N + c(w))
	\end{equation}
	we rewrite the partition function \eqref{hmm_schur_expansion} in a more convenient form \eqref{hypergeometric_tau_function_bosonic_representation} as a hypergeometric $ \tau $-function with parameters:
	\begin{equation}
		\begin{gathered}
		r(n) = N + n;\\
		\beta_{k} = \frac{1}{2} \delta_{k,2}
		\end{gathered}
	\end{equation}
	So, unrescaled Hermitian matrix model is the hypergeometric $ \tau $-function of the unrescaled KP and, in terms of this family, has the form
	\begin{equation}\label{hmm_as_hypergeometric_function_classical}
		\tau_{HMM}(\textbf{t}) \equiv Z_{N}(\textbf{t}) = \sum_{\lambda} \left( \prod_{w \in \lambda} (N + c(w)) \right) s_{\lambda}\left( \beta_{k} = \frac{1}{2} \delta_{k,2} \right) s_{\lambda} (\textbf{t})
	\end{equation}

	Note that Kazarian-Lando prescription \eqref{Kazaryan_Lando_deformation} is not applicable for this model -- the parameters $ \beta $ are different. The Pl\"{u}cker coefficients for this model satisfy the Pl\"{u}cker relations for any $ N $:
	\begin{equation}
		C_{\lambda} = \left( \prod_{w \in \lambda} (N + c(w)) \right) s_{\lambda}\left( \beta_{k} = \frac{1}{2} \delta_{k,2} \right)
	\end{equation}

	\subsubsection{HMM as a solution of $\hbar$-KP}
	It is well known how to deform Hermitian matrix model in order to distinguish different genera contributions (see, for example \cite{Alexandrov_2009}):
	\begin{equation}\label{hmm_deformed_matrix_integral}
		Z_{N}^{\hbar}(\textbf{t}) = \frac{\int DX \exp \left(-\frac{1}{2\hbar} \Tr(X^{2}) + \frac{1}{\hbar} \sum_{k=1}^{\infty} t_{k} \Tr(X^{k})\right)}{\int DX \exp \left(-\frac{1}{2\hbar} \Tr(X^{2}) \right)}
	\end{equation}
	Immediately we see that times $ t_{k} $ are rescaled correctly: $ t_{k} \rightarrow \frac{t_{k}}{\hbar} $. In order to make correct deformation we have to put $N\hbar=t_{0}$, where $t_0$ is a constant parameter and $N\rightarrow\infty$. Now if we perform character expansion of deformed model \eqref{hmm_deformed_matrix_integral} we obtain
	\begin{equation}
		Z_{N}^{\hbar}(\textbf{t}) = \sum_{\lambda} \frac{\prod_{w \in \lambda} (t_{0} + \hbar c(w)) }{\hbar^{|\lambda|/2}} s_{\lambda}\left(\beta_{n} = \frac{1}{2}\delta_{n,2}\right) s_{\lambda}\left(\frac{\textbf{t}}{\hbar}\right).
	\end{equation}
	Note that $ \hbar^{-|\lambda|/2} $ can be introduced into the Schur polynomial
	\begin{equation}
		\tau_{HMM}^{\hbar}(\textbf{t}) \equiv Z_{N}^{\hbar}(\textbf{t}) = \sum_{\lambda} \left( \prod_{w \in \lambda} (t_{0} + \hbar c(w)) \right) s_{\lambda}\left( \beta_{k} = \frac{1}{2\hbar} \delta_{k,2} \right) s_{\lambda} \left( \frac{\textbf{t}}{\hbar} \right)
	\end{equation}
	with deformed Pl\"{u}cker coefficients
	\begin{equation}
		C_{\lambda}^{\hbar} = \left( \prod_{w \in \lambda} (t_{0} + \hbar c(w)) \right) s_{\lambda}\left( \beta_{k} = \frac{1}{2\hbar} \delta_{k,2} \right).
	\end{equation} 

	It is important to note that, as in the case of Hurwitz numbers, $ \tau_{HMM}^{\hbar}(\textbf{t}) $ decomposes both into connected graphs and disconnected ones, therefore the logarithm of a $ \tau $-function has "correct" $ \; $ quasi-classical behaviour, which is expanded only in connected graphs
	\begin{equation}
		F_{HMM}^{\hbar}(\textbf{t}) = \sum_{g=0}^{\infty} \hbar^{2g} F_{HMM}^{g}(\textbf{t}).
	\end{equation}
	As a result we can rewrite deformation as:
	\begin{equation}\label{hmm_correct_deformation}
		\boxed{
			\begin{gathered}
			r(n) = (t_{0} + n) \rightarrow r(\hbar n) = (t_{0} + \hbar n)\\
			\beta_{2} \rightarrow \frac{\beta_{2}}{\hbar}\\
			t_{k} \rightarrow \frac{t_{k}}{\hbar}
			\end{gathered}}
	\end{equation}

	Now let us show that deformed HMM is a solution of $\hbar$-KP. We will do it in the same way as for Hurwitz numbers. Following Proposition 2, we have to show that deformed coefficients $C_{\lambda}^{\hbar}$ satisfy the classical Pl\"{u}cker relations. For this let us rewrite $ C_{\lambda}^{\hbar} $ as:
	\begin{equation}
		C_{\lambda}^{\hbar} = \frac{1}{\hbar^{|\lambda|/2}} \left( \prod_{w \in \lambda} (t_{0} + \hbar c(w)) \right) s_{\lambda}\left( \beta_{k} = \frac{1}{2} \delta_{k,2} \right) = \hbar^{|\lambda|/2} \left( \prod_{w \in \lambda} \left[ \frac{t_{0}}{\hbar} + c(w) \right] \right) s_{\lambda}\left( \beta_{k} = \frac{1}{2} \delta_{k,2} \right)
	\end{equation} 
	Similarly to Hurwitz numbers, $\hbar^{|\lambda|/2}$ factor does not change the Pl\"{u}cker relations (because of homogeneity), which hold for the rest of the expression. Therefore, deformed Hermitian matrix model is a solution of $\hbar$-KP.

	As HMM $\tau$-function is function of hypergeometric type, one can use \eqref{hmm_correct_deformation} to write matrix in fermionic representation:
	\begin{equation}
		A_{HMM}^{\hbar} = \frac{1}{\hbar} \sum_{n \in \mathbb{Z}} (t_{0} + \hbar n)(t_{0} + \hbar (n-1)) \colon \psi_{n} \psi_{n-2}^{*} \colon = \frac{1}{\hbar} \oint \frac{dz}{2 \pi i} \colon \left[ \left(\frac{1}{z} (t_{0} + \hbar D)\right)^{2} \psi(z) \right] \cdot \psi^{*}(z) \colon
	\end{equation}
	which is the same as Takasaki-Takebe deformation \eqref{Takasaki_Takebe_deformation}. Yet again we see that $ a_{i,j}(\hbar) $ do not, in fact, depend on $\hbar$.

	\subsection{Kontsevich model}
	In this section we discuss the Kontsevich $\tau$-function and show that a rescaled $\tau$-function is the solution of $\hbar$-KP. Here we do not consider classical case and start with already deformed function and consider it in bosonic representation.
 	Kontsevich model is a generating function for the intersection numbers of Chern classes on compactified moduli spaces $\overline{\mathcal{M}}_{g;n}$ of complex curves of genus $ g $ with $ n $ marked points. Intersection numbers of Chern classes
	\begin{equation}
		\int_{\overline{\mathcal{M}}_{g;n}} \psi_{1}^{m_{1}} \psi_{2}^{m_{2}} \dots \psi_{n}^{m_{n}} = \langle \tau_{m_{1}} \tau_{m_{2}} \dots \tau_{m_{n}} \rangle
	\end{equation}
	are rational numbers, which are not equal to zero only if
	\begin{equation}\label{condition_Kontsevich}
		\sum_{i=1}^{n} (m_{i} - 1) = 3g - 3.
	\end{equation}
	Let us define generating function with parameter $\hbar$ enumerating contributions of different genera \cite{Alexandrov_2013}:
	\begin{equation}\label{Kontsevich_F_function}
		\begin{gathered}
		F_{K}^{\hbar}(T_{k}) = \hbar^{2} \left\langle \exp\left( \sum_{m=0}^{\infty}(2m+1)!! \hbar^{\frac{2(m-1)}{3}} T_{m} \tau_{m} \right) \right\rangle = \sum_{g=0}^{\infty} \hbar^{2g} F_{K}^{g}(T_{k}) 
		\end{gathered}
	\end{equation}

	It is known \cite{kontsevich1992intersection} that  $ Z_{K}^{\hbar}(T_{k}) = \exp\left( \frac{1}{\hbar^{2}} F_{K}^{\hbar}(T_{k}) \right) $ defined by the Kontsevich matrix integral
	\begin{equation}\label{Kontsevich_integral_form}
		Z_{K}^{\hbar}(T_{k}) = \exp\left( \frac{1}{\hbar^{2}} F_{K}^{\hbar}(T_{k}) \right) = \frac{\int DX \exp\left( \frac{1}{\hbar} \Tr\left( i\frac{X^{3}}{3!} + \frac{\Lambda X^{2}}{2} \right) \right)}{\int DX \exp \left( \frac{1}{\hbar} \Tr \frac{\Lambda X^{2}}{2} \right)}
	\end{equation}
	where integration is taken over hermitian matrices $ X $. Furthermore, it is a $\tau$-function of the KdV hierarchy for $\hbar=1$. From the point of view of KP it depends only on odd times $t_k$, that in terms of matrix model have the form
	\begin{equation}
		T_{k} = t_{2k+1} = \frac{\hbar}{2k+1} \Tr \Lambda^{-2k-1}
	\end{equation}
	From \eqref{Kontsevich_F_function} we can see that genus expansion is obtained by rescaling "times":
	\begin{equation}\label{known_deformation}
		t_{2k+1} \rightarrow \hbar^{\frac{2(k-1)}{3}} t_{2k+1}
	\end{equation}

	Let us show that deformation \eqref{known_deformation} is equivalent to:
	\begin{equation}\label{kontsevich_correct_deformation}
		\boxed{
			\begin{gathered}
			t_{k} \rightarrow \frac{t_{k}}{\hbar}\\
			C_{\lambda} \rightarrow C_{\lambda}^{\hbar} = C_{\lambda} \hbar^{\frac{|\lambda|}{3}}
			\end{gathered}}
	\end{equation}
	Suppose we decompose a $ \tau $-function with deformation \eqref{kontsevich_correct_deformation} over Schur polynomials. Then an arbitrary monomial in $ \tau_{K}^{\hbar}(\textbf{t}) $ has the form: 
	\begin{equation}
		C_{\lambda}^{\hbar} \prod_{i=1}^{k} \frac{t_{2m_{i}+1}}{\hbar} = C_{\lambda} \hbar^{\frac{|\lambda|}{3} - k} \prod_{i=1}^{k} t_{2m_{i}+1}
	\end{equation}
	now, using simple relation
	\begin{equation}
		|\lambda| = \sum_{i=1}^{k} (2m_{i} + 1) = \left(\sum_{i=1}^{k} 2(m_{i} - 1)\right) + 3k
	\end{equation}
	we obtain
	\begin{equation}
		C_{\lambda} \hbar^{\frac{|\lambda|}{3} - k} \prod_{i=1}^{k} t_{2m_{i}+1} = C_{\lambda} \hbar^{\frac{1}{3} \left(\sum_{i=1}^{k} 2(m_{i} - 1)\right)} \prod_{i=1}^{k} t_{2m_{i}+1} = C_{\lambda} \prod_{i=1}^{k} \hbar^{\frac{2(m_{i}-1)}{3}} t_{2m_{i}+1}.
	\end{equation}
	Using the same argument about homogeneity, we obtain that coefficients $C_{\lambda}^{\hbar}$ satisfy the Pl\"{u}cker relations since non-deformed coefficients $C_{\lambda}$ satisfy them in the first place. As a result, $ \tau_{K}^{\hbar}(\textbf{t}) \in \hbar $-KP.

	\subsection{Brezin-Gross-Witten model} 
	In this section we discuss Brezin-Gross-Witten matrix model and its deformation. We consider the model in two phases and show that in both of them the deformed model is a solution of $\hbar$-KP.
	
	\subsubsection{Classical BGW model}
	Partition function of the model is defined as follows \cite{brezin1980external, gross1980possible}
	\begin{equation}\label{BGW}
		Z_{BGW}(J,J^{+})=\frac{1}{V_N}\int_{N\times N} DU\exp( \Tr(J^{+}U+JU^{+}))\:,
	\end{equation}
	where we integrate over unitary matrices $ N \times N $ with the Haar measure $ DU $ and $ V_{N} = \int_{N \times N} DU $ is the volume of the unitary group.

	Since the Haar measure is invariant with respect to group action, $ Z_{BGW} (J, J^{+}) $ depends only on $ N $ parameters, eigenvalues of the matrix $ JJ^{+} $. Depending on the choice of variables $ t_k $, we consider two phases \cite{mironov1996unitary}:
	\begin{equation}
		t_k=\frac{1}{k}Tr(JJ^{+})^k \: \text{-- character phase}
	\end{equation}
	\begin{equation}\label{Kontsevich_times}
		t_k=-\frac{1}{2k-1}Tr(JJ^{+})^{-k+\frac{1}{2}}\: \text{-- Kontsevich phase}
	\end{equation}
	We denote the partition function as a series in $ t_k $, in the character phase as $ Z^{+}_{BGW} $, in the Kontsevich phase as $ Z^{-}_{BGW} $. BGW model in character phase is a KP $\tau$-function and in Kontsevich phase a KdV $\tau$-function, thus, a KP $\tau$-function too.

	The character phase has a simple expansion over Schur polynomials \cite{morozov2010unitary}:
	\begin{equation}
		Z^{+}_{BGW}(J,J^{+})=\sum_{\lambda}\frac{d^2_{\lambda}}{D_{\lambda}}\chi_{\lambda}(JJ^{+})=\sum_{\lambda}\frac{d^2_{\lambda}}{D_{\lambda}}s_{\lambda} \left(t_k= \frac{Tr(JJ^{+})^k}{k} \right)    
	\end{equation}
	Using relation \eqref{dimensions_ratio} we see that BGW model in the character phase is a hypergeometric $ \tau $-function with parameters $\beta_{n} = \delta_{n,1} $:
	\begin{equation}
		Z^{+}_{BGW}=\sum_{\lambda} \left(\prod_{\omega\in\lambda}\frac{1}{N+c(\omega)}\right)s_{\lambda}(\beta_{n} = \delta_{n,1})s_{\lambda}(t)
	\end{equation}
	
	\subsubsection{BGW as a solution of $\hbar$-KP}
	Deformation of the model in both phases is given by introduction of $ 1/\hbar $ into the exponent under the matrix integral
	\begin{equation}\label{BGW_deformed}
		Z_{\hbar BGW} = \frac{1}{V_N}\int_{N\times N} DU\exp\left(\frac{1}{\hbar} \Tr(J^{+}U+JU^{+})\right),
	\end{equation}
	with additional requirement $ N \hbar = t_{0} $, 
	where $ t_{0} $ is a constant parameter and $ N \to \infty $, which is common for matrix models. 
	
	Let us first consider the character phase. Time variables are rescaled as usual
	\begin{equation}\label{BGW_character_phase_times_deformed}
		t_k=\frac{\hbar}{k}Tr(JJ^{+})^k.
	\end{equation}
	If we now perform character expansion for the deformed model \eqref{BGW_deformed} with times \eqref{BGW_character_phase_times_deformed} we obtain
	\begin{equation}
		Z^{+}_{\hbar BGW}=\sum_{\lambda} \left(\prod_{\omega\in\lambda}\frac{1}{t_{0} + \hbar c(\omega)}\right)s_{\lambda}\left( \beta_{n} = \frac{1}{\hbar}\delta_{n,1} \right) s_{\lambda}\left( \frac{\textbf{t}}{\hbar} \right).
	\end{equation}
	Note that it is deformed in accordance with prescription \eqref{hurwitz_correct_deformation}. With the help of similar arguments as for Hurwitz numbers, $ Z_{\hbar BGW}^{+} $ is a solution of $\hbar$-KP.

	The first few terms of the genus expansion of $F_{\hbar BGW}^{+}=\hbar^2\log(Z^{+}_{\hbar BGW})$ are:
	\begin{equation}
		F_{\hbar BGW}^{+}=\left(\frac{t_1}{t_{0}} + \frac{t_1^2}{2t_{0}^4} - \frac{t_2}{t_{0}^3} + \dots\right) 
		+\hbar^2\left(\frac{t_1^2}{2t_{0}^6} - \frac{t_2}{t_{0}^5} + \dots\right) + \hbar^4\left(\frac{t_1^2}{2t_{0}^8} - \frac{t_2}{t_{0}^7} + \dots\right) + \dots
	\end{equation}

	In Kontsevich phase things are a little different since $Z^{-}_{\hbar BGW}$ is not a hypergeometric $\tau$-function. The same deformed matrix integral \eqref{BGW_deformed} gives genus expansion for this phase. Deformation in this phase can be written only in terms of rescaling of times \cite{Alexandrov_2018}:
	\begin{equation}\label{known_deformation_BGW}
		t_k\rightarrow t_k \hbar^{k-1}
	\end{equation}
	Therefore, if the initial partition function $ Z^{-}_{BGW} $ has expansion over Schur polynomials of the form
	\begin{equation}
		Z_{BGW}^{-}=\sum_{\lambda} C_{\lambda}s_{\lambda}(\textbf{t})\:,
	\end{equation}
	then we can use the same arguments as for Kontsevich model (formula \eqref{kontsevich_correct_deformation}) to state that rescaling \eqref{known_deformation_BGW} is equivalent to the following deformation
	\begin{equation}
		\boxed{
			\begin{gathered}
			t_{k} \rightarrow \frac{t_{k}}{\hbar},\\
			C_{\lambda} \rightarrow C_{\lambda}^{\hbar} = C_{\lambda}\hbar^{|\lambda|}
			\end{gathered}
		}
	\end{equation}
	Now, using our usual argument about homogeneity of the Pl\"{u}cker relations, we obtain that coefficients $C_{\lambda}^{\hbar}=\hbar^{|\lambda|}C_{\lambda}$ satisfy the classical Pl\"{u}cker relations. We conclude that $Z_{\hbar BGW}^{-}$ is the $\tau$-function of $\hbar$-KP.

	Finally, it is interesting to look at the few first terms of genus expansion:
	\begin{equation}
		F^{-}_{\hbar BGW}=\hbar^2\log(Z^{-}_{\hbar BGW})=0-\hbar^2\frac{1}{2}\log(1-\frac{t_1}{2})+\dots
	\end{equation}
	Note that it has zero quasi-classical limit which is the same as for the trivial $\tau$-function $\tau\equiv 1$.

	\section{Generalizing deformation}
	As we have seen on examples of Hurwitz numbers, Hermitian matrix model and BGW model in character phase, they all are deformed similarly because they belong to one set of hypergeometric solutions (despite the fact that the set of parameters $\beta_{k}$ is different). Let us generalize deformation of Orlov-Scherbin family to the entire hypergeometric family:
	\begin{equation}\label{deformation_rule_hypergeometric_tau_function}
	\boxed{
		\begin{gathered}
		r(n) \rightarrow r(\hbar n)\\
		\beta_{n} \rightarrow \frac{\beta_{n}}{\hbar}\\
		t_{n} \rightarrow \frac{t_{n}}{\hbar}\\
		\end{gathered}
	}
	\end{equation}
	Deformation of Hurwitz numbers \eqref{hurwitz_correct_deformation} and Hermitian matrix model \eqref{hmm_correct_deformation} are special cases of deformation \eqref{deformation_rule_hypergeometric_tau_function}. Such a deformation of $\tau$-functions in the language of fermionic represenation \eqref{Takasaki_Takebe_deformation} is the following: 
	\begin{itemize}
		\item $ r(n) \rightarrow r(\hbar n) $ corresponds to the rescaling of the derivative $ \frac{d}{dz} \rightarrow \hbar \frac{d}{dz} $,
		\item $ \beta_{n} \rightarrow \frac{\beta_{n}}{\hbar} $ correspond to the factor $ \frac{1}{\hbar} $ in front of integral over fermionic fields.
		\item $ t_{n} \rightarrow \frac{t_{n}}{\hbar} $ corresponds to the change of variables $ H(\textbf{t}) \rightarrow \frac{H(\textbf{t})}{\hbar} $.
	\end{itemize}
	Therefore, in the fermionic language $\hbar$-deformation is universal and does not belong on a particular model being considered. Using Proposition 1 we conclude that such deformed $\tau$-functions have "good" quasi-classical behaviour, i.e. function $ F^{\hbar}(\textbf{t}) = \hbar^{2} \log(\tau^{\hbar}(\textbf{t})) $ has only non-negative powers of $ \hbar $. As it was mentioned in section 5.1.2, if parameters $ \beta = \{1,0,0, \dots\} $ then such deformation leads to decomposition over even powers of $ \hbar $. In the case of arbitrary set of parameters $ \beta $ it is unknown if there will be only even powers of $ \hbar $ or not.
	
	\textbf{Proposition 5.} \textit{Deformation \eqref{deformation_rule_hypergeometric_tau_function} of hypergeometric $\tau$-function is $\tau$-function of $\hbar$-KP}
	
	\textbf{Proof:} Schur polynomials $ s_{\lambda}(\beta) $ satisfy the Pl\"{u}cker relations for any $ \beta $ \cite{miwa2000solitons}, one can easily check first of them \eqref{plucker_relations}:
	\begin{equation}
		s_{[2,2]}(\beta) s_{[\varnothing]}(\beta) - s_{[2,1]}(\beta) s_{[1]}(\beta) + s_{[2]}(\beta) s_{[1,1]}(\beta) = 0.
	\end{equation}
	$ s_{\lambda}(\beta/\hbar) $ then satisfy the Pl\"{u}cker relations too. In the classical case multiplication of $ s_{\lambda}(\beta) $ on $ r_{\lambda} = \prod_{w \in \lambda} r(c(w)) $ does not violate the Pl\"{u}cker relations because all terms contain the same set of contents $ c(w) $:
	\begin{equation}
		\begin{gathered}
		r(0)^{2}r(1)r(-1)s_{[2,2]}(\beta) s_{[\varnothing]}(\beta) - r(0)r(1)r(-1)s_{[2,1]}(\beta) r(0)s_{[1]}(\beta) + r(0)r(1)s_{[2]}(\beta) r(0)r(-1)s_{[1,1]}(\beta) = \\
		= r(0)^{2}r(1)r(-1) \left[ s_{[2,2]}(\beta) s_{[\varnothing]}(\beta) - s_{[2,1]}(\beta) s_{[1]}(\beta) + s_{[2]}(\beta) s_{[1,1]}(\beta) \right] = 0.
		\end{gathered}
	\end{equation}
	Deformed case is similar, $ C_{\lambda}^{\hbar} = \prod_{w \in \lambda} r(\hbar c(w)) s_{\lambda}(\beta/\hbar) $ again satisfy the Pl\"{u}cker relations, therefore, deformed hypergeometric $\tau$-functions are solutions of $\hbar$-KP. $ \square $ 
	
	Thus, there is entire set of $\hbar$-hypergeometric $\tau$-functions that are $\hbar$-KP solutions with good quasi-classical behaviour. As an illustration of this approach to deformation let us consider an example of "trivial" $\tau$-function
	\begin{equation}
		\tau(\textbf{t};\beta) = \sum_{\lambda} s_{\lambda}(\beta) s_{\lambda}(t) = \exp \left( \sum_{n=0}^{\infty} n \beta_{n} t_{n} \right)
	\end{equation}
	which is a hypergeometric $\tau$-function with $ r(n) \equiv 1 $ and arbitrary parameters $\beta$. In the last formula Cauchy-Littlewood identity \eqref{Cauchy-Littlewood_identity} was used. If we apply now deformation \eqref{deformation_rule_hypergeometric_tau_function} to this function, we obtain
	\begin{equation}
		\tau^{\hbar}(\textbf{t};\beta) = \sum_{\lambda} s_{\lambda}\left( \frac{\beta}{\hbar} \right) s_{\lambda}\left( \frac{t}{\hbar} \right) = \exp \left( \sum_{n=0}^{\infty} \frac{n \beta_{n} t_{n}}{\hbar^{2}} \right)
	\end{equation}
	Free energy is provided with good quasi-classical behaviour (the entire function is of genus zero):
	\begin{equation}
		F^{\hbar}(\textbf{t}; \beta) = \hbar^{2} \log(\tau^{\hbar}(\textbf{t},\beta)) = \hbar^{2} \sum_{n=0}^{\infty} \frac{n \beta_{n} t_{n}}{\hbar^{2}} \equiv F^{g=0}(\textbf{t}, \beta)
	\end{equation}
	So, everything fits together nicely.
	
	
	\section{$\hbar$-KP and the Pl\"{u}cker relations}
	In Section 4.2, $ \tau $-functions of $ \hbar $-KP with two parameters $ \hbar $ and $ x $ were introduced. These parameters usually are not presented in the classical KP hierarchy. Let us try to understand how the various relations from the theory of $ \hbar $-KP are related to the already known formulas.
		
	\textbf{Proposition 6.} \textit{Determinant relations \eqref{coefficient_determinant_relations} on functions $ C_{\lambda}^{\hbar}(x) $ are linear combination of the classical Pl\"{u}cker relations}
		
		\textbf{Proof:} The $ x $ parameter is added when the first variable $ t_{1} \rightarrow t_{1} + x $ is shifted. On the one hand, it is known that we should get:
		\begin{equation}\label{shifted_tau_function}
		\tau^{\hbar}(t_1+x,t_2,...)=\sum\limits_{\lambda} C_{\lambda}^{\hbar}(x)s_{\lambda} \left( \frac{t_1}{\hbar}, \frac{t_2}{\hbar},.. \right)
		\end{equation}
		on the other hand, one can explicitly shift the first variable
		\begin{equation}
		\tau^{\hbar}(t_1+x,t_2,...)= \sum\limits_{\lambda} C_{\lambda}^{\hbar} s_{\lambda} \left( \frac{t_1 + x}{\hbar}, \frac{t_2}{\hbar},.. \right)
		\end{equation}
		Shifted Schur polynomials can be decomposed over basis of usual Schur polynomials as follows:	\begin{equation}\label{shifted_schur_polynomials}
		s_{\lambda}(t_{1} + x, t_{2}, \dots) = \sum\limits_{\{\nu|\nu \subset \lambda\}} D(\lambda, \mu) \frac{x^{|\lambda| - |\mu|}}{(|\lambda| - |\mu|)} s_{\mu}(\textbf{t})
		\end{equation}
		where coefficients $ D(\lambda,\mu) $ have combinatorial description: it is the number of ways to get Young diagram $\lambda$ from $\mu$ by adding $|\lambda|-|\mu|$ boxes in sequence in a consistent way so that we get a Young diagram at every step:
		\begin{equation}\label{combinatorial_coefficients}
		D(\lambda,\mu) = s_{\lambda}(\tilde\partial)(s_{[1]}(t))^{|\lambda|-|\mu|}s_{\mu}(t)\Big|_{\textbf{t}=0}
		\end{equation}
		Then from \eqref{shifted_schur_polynomials} we can find an explicit expression for functions $ C_{\lambda}^{\hbar}(x) $:
		\begin{equation}\label{hurwitz_coefficients}
		C_{\lambda}^{\hbar}(x) = \sum\limits_{\{\mu: \lambda \subset \mu\}} e^{\hbar u c(\lambda)} \frac{s_{\mu}(\beta_{n} = \delta_{n,1})}{\hbar^{|\mu|}} D(\mu,\lambda) \frac{(x/\hbar)^{|\mu| - |\lambda|}}{(|\mu| - |\lambda|)!}
		\end{equation} 
		The main property of coefficients $ D(\mu,\nu) $ is:
		\begin{equation}\label{combinatorial_coefficients_property}
		D(\mu,\nu) = \sum_{\{\lambda:|\lambda|=k\}} D(\mu,\lambda) D(\lambda,\nu),\; \forall k:\: |\nu|\leq k\leq |\mu|
		\end{equation}
		which follows from their combinatorial description.
		Here we put $C(\mu,\mu)=1$ and $C(\mu,\nu)=0$ if $\nu\not\subset\mu$. Now, using this property, we obtain the rule of differentiation of $ C_{\mu}^{\hbar}(x) $:
		\begin{equation}\label{hurwitz_coefficients_property}
		\hbar\partial C_{\mu}^{\hbar}(x)=\sum\limits_{\lambda=\mu+[1]} C_{\lambda}^{\hbar}(x)
		\end{equation}
		where we sum over all Young diagrams obtained from $\mu$ by adding one box. It is more convenient for further calculation to parametrize $ C_{\mu} $ by Maya diagrams instead of Young diagrams. In this terms the last property can be rewritten in the form:
		\begin{equation}\label{hurwitz_coefficients_property_maya_diagrams}
		\hbar\partial C_{\{m_1,m_2,...\}}^{\hbar}(x) = \sum\limits_{k=1}^{\infty} C_{\{m_1,...,m_k-1,...\}}^{\hbar}(x)
		\end{equation}
		Using formula \eqref{hurwitz_coefficients_property_maya_diagrams} for derivatives of $ C_{\mu}^{\hbar}(x) $, the determinant relations \eqref{coefficient_determinant_relations} for $ \hbar $-KP can be reduced to the Pl\"{u}cker relations as follows. First of all, we expand the determinant corresponding to the diagram $ \lambda $ in the last column. In terms of Maya diagrams:
		\begin{equation}\label{determinant_decomposition}
		\det_{\lambda} = (-1)^{l(\lambda)} \sum_{n=1}^{l(\lambda)} (-1)^{n} C^{\hbar}_{\{m_1, \dots, \hat{m}_{n}, \dots, m_{l(\lambda)}, l(\lambda) - \frac{1}{2}, \dots\}}(x) \left( \sum_{k=0}^{l(\lambda) -1} (-1)^{k} \binom{j-1}{k} (\hbar \partial)^{k} C^{\hbar}_{\{m_{n} + 1 - l(\lambda) + k, \frac{3}{2}, \dots\}}(x) \right)
		\end{equation}
		Here the coefficient in brackets is the minor of the matrix of size $ (l (\lambda) - 1) $, which was obtained by deleting the last column and some row. Then
		\begin{multline}\label{has_to_be_zero}
			C^{\hbar}_{\lambda}(x) C^{\hbar}_{\varnothing}(x) - \det_{\lambda} = (-1)^{l(\lambda) + 1} \sum_{n=1}^{l(\lambda) + 1} (-1)^{n} C^{\hbar}_{\{m_1, \dots, \hat{m}_{n}, \dots, m_{l(\lambda)}, l(\lambda) - \frac{1}{2}, \dots\}}(x) \cdot \\
			\cdot \left( \sum_{k=0}^{l(\lambda) -1} (-1)^{k} \binom{j-1}{k} (\hbar \partial)^{k} C^{\hbar}_{\{m_{n} + 1 - l(\lambda) + k, \frac{3}{2}, \dots\}}(x) \right)
		\end{multline}
		
		The Pl\"{u}cker relations in terms of Maya diagrams can be expressed as follows:
		\begin{equation}\label{Plucker_equations_Maya_diagrams}
		\sum_{k=1}^{\infty} C^{\hbar}_{\{m_1, \dots, \hat{m}_{k}, \dots \}}(x) C^{\hbar}_{\{m_{k}, n_{1}, \dots \}}(x) = 0
		\end{equation}
		Now let us introduce operators $ i_{n} $ and $ P $:
		\begin{equation}
		i_{m_k} C^{\hbar}_{\{n_{1}, \dots \}}(x) = C^{\hbar}_{\{ m_{k}, n_{1}, \dots \}}(x)
		\end{equation}
		\begin{equation}
		P = \sum_{k=1}^{\infty} (-1)^{k} C^{\hbar}_{\{m_1, \dots, \hat{m}_{k}, \dots \}}(x) \; i_{m_k}
		\end{equation}
		From \eqref{Plucker_equations_Maya_diagrams} it is clear, that $ P $ gives LHS of the Pl\"{u}cker relations acting on $ C_{\lambda}^{\hbar}(x) $. Commutation relation between these operators is
		\begin{equation}\label{property_of_i_k}
		[i_{m_k}, \hbar \partial_{x}] = -i_{m_{k} - 1}
		\end{equation}
		Using it one can calculate:
		\begin{equation}
		i_{m_k} (\hbar \partial)^{l} = \sum_{p = 0}^{k} (-1)^{p} \binom{k}{p} (\hbar \partial)^{k-p} i_{m_{k} - p}
		\end{equation}
		and
		\begin{equation}
		P (\hbar \partial)^{l-1} = (-1)^{l-1} \sum_{n=1}^{\infty} (-1)^{n} C^{\hbar}_{\{m_1, \dots, \hat{m}_{n}, \dots \}}(x) \left( \sum_{k = 0}^{l - 1} (-1)^{k} \binom{l-1}{k} (\hbar \partial)^{k-p} i_{m_{n} - l + 1 + k} \right)
		\end{equation}
		Acting by the last operator on $ C^{\hbar}_{\{\frac{3}{2}, \frac{5}{2}, \dots \}}(x) $ and using the property \eqref{hurwitz_coefficients_property_maya_diagrams}, we get exactly the RHS of \eqref{has_to_be_zero}. This means that we have reduced the determinant formula to a linear combination of the Pl\"{u}cker relations. $ \square $

			\section{Conclusion}
			In conclusion we list the main results obtained in the paper:
		\begin{itemize}
			
			\item 
				We have considered several examples of $ \tau $-functions that have natural genus expansion, that is, introduction of parameter $\hbar$ distinguishes surfaces of different genera. We have shown that deformation of any considered example is the $ \tau $-function of $ \hbar $-KP. Thus, on these examples we have explicitly shown that natural genus expansion coincides with expansion in $\hbar$ for $\hbar$-KP.
			
			\item 
				We have shown that deformation considered in \cite{kazarian2015combinatorial} is a particular case of Takasaki-Takebe approach \cite{TAKASAKI_1995}. $\hbar$-formulation of \cite{Natanzon_2016} with parameter $ x=0 $ again coincides with Takasaki-Takebe one.
			
			\item 
				Some of mentioned examples are known to be hypergeometric $ \tau $-functions. It was possible to generalize the deformation described in \cite{kazarian2015combinatorial} to the entire hypergeometric family of solutions. Such $\tau$-functions are solutions of $\hbar$-KP with good quasi-classical limit (Proposition 5).
			
			
			\item 
				It was explicitly shown how the determinant relations \eqref{coefficient_determinant_relations} for formal solutions of $ \hbar $-KP are related to the classical the Pl\"{u}cker relations (Proposition 6).
		\end{itemize}
		
			Let us formulate some questions with which further research on the $ \hbar $-KP may be related:
		\begin{itemize}
			\item 
				In each of the considered examples, there is a geometrical interpretation of the genus decomposition, and this decomposition can be obtained using topological recursion. How does this procedure look in terms of $ \hbar $-KP? That is, is there a combinatorial way to recover the higher genera for some class of $ \tau $ -functions, for example, hypergeometric ones?
			
			\item 
				Another question may be related to the structure of formal solutions of the $ \hbar $-KP described in the work of Natanzon-Zabrodin. Is it possible, for example, to obtain the determinant relations \eqref{coefficient_determinant_relations} as Ward identities in one of the matrix models?
		\end{itemize}

	\section{Acknowledgements}
	This work was funded by the Russian Science Foundation (Grant No.20-71-10073). We are grateful to Alexander Alexandrov, Andrey Mironov and Takashi Takebe for very useful discussions and remarks. Our special acknowledgement is to Sergey Natanzon for a formulation of the problem and to Anton Zabrodin for pointing out several examples. 
	
	
	\bibliographystyle{ieeetr}
	\bibliography{h_KP_deformation}

\begin{thebibliography}{10}

\bibitem{TAKASAKI_1995}
K.~Takasaki and T.~Takebe, ``Integrable hierarchies and dispersionless limit,''
  {\em Reviews in Mathematical Physics}, vol.~07, p.~743–808, Jul 1995.

\bibitem{Natanzon_2016}
S.~M. Natanzon and A.~V. Zabrodin, ``Formal solutions to the {KP} hierarchy,''
  {\em Journal of Physics A: Mathematical and Theoretical}, vol.~49, p.~145206,
  Feb 2016.

\bibitem{kazarian2015combinatorial}
M.~E. Kazarian and S.~K. Lando, ``Combinatorial solutions to integrable
  hierarchies,'' {\em Russian Mathematical Surveys}, vol.~70, no.~3, p.~453,
  2015.

\bibitem{saad2019jt}
P.~Saad, S.~H. Shenker, and D.~Stanford, ``{J}{T} gravity as a matrix
  integral,'' {\em arXiv preprint arXiv:1903.11115}, 2019.

\bibitem{witten2020volumes}
E.~Witten, ``Volumes and random matrices,'' {\em arXiv preprint
  arXiv:2004.05183}, 2020.

\bibitem{witten2020matrix}
E.~Witten, ``Matrix {Models} and {Deformations} of {J}{T} {Gravity},'' {\em
  arXiv preprint arXiv:2006.13414}, 2020.

\bibitem{Jensen_2016}
K.~Jensen, ``Chaos in ${A}d{S}_2$ {Holography},'' {\em Physical Review
  Letters}, vol.~117, Sep 2016.

\bibitem{Maldacena_2016_1}
J.~Maldacena, D.~Stanford, and Z.~Yang, ``Conformal symmetry and its breaking
  in two-dimensional nearly anti-de {Sitter} space,'' {\em Progress of
  Theoretical and Experimental Physics}, vol.~2016, p.~12C104, Nov 2016.

\bibitem{Engels_y_2016}
J.~Engelsöy, T.~G. Mertens, and H.~Verlinde, ``An investigation of
  ${A}d{S}_{2}$ backreaction and holography,'' {\em Journal of High Energy
  Physics}, vol.~2016, Jul 2016.

\bibitem{Kitaev_2018}
A.~Kitaev and S.~J. Suh, ``The soft mode in the {Sachdev}-{Ye}-{Kitaev} model
  and its gravity dual,'' {\em Journal of High Energy Physics}, vol.~2018, May
  2018.

\bibitem{Maldacena_2016}
J.~Maldacena and D.~Stanford, ``Remarks on the {Sachdev}-{Ye}-{Kitaev} model,''
  {\em Physical Review D}, vol.~94, Nov 2016.

\bibitem{Witten_2019}
E.~Witten, ``An {S}{Y}{K}-like model without disorder,'' {\em Journal of
  Physics A: Mathematical and Theoretical}, vol.~52, p.~474002, Oct 2019.

\bibitem{Gurau_2017}
R.~Gurau, ``The complete 1/{N} expansion of a {S}{Y}{K}–like tensor model,''
  {\em Nuclear Physics B}, vol.~916, p.~386–401, Mar 2017.

\bibitem{amburg2020correspondence}
N.~Amburg, H.~Itoyama, A.~Mironov, A.~Morozov, D.~Vasiliev, and R.~Yoshioka,
  ``Correspondence between {Feynman} diagrams and operators in quantum field
  theory that emerges from tensor model,'' {\em The European Physical Journal
  C}, vol.~80, pp.~1--5, 2020.

\bibitem{itoyama2020complete}
H.~Itoyama, A.~Mironov, and A.~Morozov, ``Complete solution to {Gaussian}
  tensor model and its integrable properties,'' {\em Physics Letters B},
  vol.~802, p.~135237, 2020.

\bibitem{klebanov2017tasi}
I.~R. Klebanov, F.~Popov, and G.~Tarnopolsky, ``T{A}{S}{I} lectures on large
  {N} tensor models,'' {\em Proceedings of Science}, vol.~305, 2017.

\bibitem{Lodin_2019}
R.~Lodin, A.~Popolitov, S.~Shakirov, and M.~Zabzine, ``Solving q-{Virasoro}
  constraints,'' {\em Letters in Mathematical Physics}, vol.~110, p.~179–210,
  Sep 2019.

\bibitem{Cassia_2019}
L.~Cassia, R.~Lodin, A.~Popolitov, and M.~Zabzine, ``Exact {S}{U}{S}{Y}
  {Wilson} loops on ${S}^{3}$ from q-{Virasoro} constraints,'' {\em Journal of
  High Energy Physics}, vol.~2019, Dec 2019.

\bibitem{chekhov2006free}
L.~Chekhov, B.~Eynard, and N.~Orantin, ``Free energy topological expansion for
  the 2-matrix model,'' {\em Journal of High Energy Physics}, vol.~2006,
  no.~12, p.~053, 2006.

\bibitem{chekhov2006hermitian}
L.~Chekhov and B.~Eynard, ``Hermitian matrix model free energy: {Feynman} graph
  technique for all genera,'' {\em Journal of High Energy Physics}, vol.~2006,
  no.~03, p.~014, 2006.

\bibitem{eynard2007invariants}
B.~Eynard and N.~Orantin, ``Invariants of algebraic curves and topological
  expansion,'' {\em arXiv preprint math-ph/0702045}, 2007.

\bibitem{eynard2005topological}
B.~Eynard, ``Topological expansion for the 1-hermitian matrix model correlation
  functions,'' {\em Journal of High Energy Physics}, vol.~2004, no.~11, p.~031,
  2005.

\bibitem{alexandrov2004partition}
A.~Alexandrov, A.~Morozov, and A.~Mironov, ``Partition functions of matrix
  models: first special functions of string theory,'' {\em International
  Journal of Modern Physics A}, vol.~19, no.~24, pp.~4127--4163, 2004.

\bibitem{alexandrov2007m}
A.~S. Alexandrov, A.~D. Mironov, and A.~Y. Morozov, ``M-theory of matrix
  models,'' {\em Theoretical and Mathematical Physics}, vol.~150, no.~2,
  pp.~153--164, 2007.

\bibitem{alexandrov2007instantons}
A.~Alexandrov, A.~Mironov, and A.~Morozov, ``Instantons and merons in matrix
  models,'' {\em Physica D: Nonlinear Phenomena}, vol.~235, no.~1-2,
  pp.~126--167, 2007.

\bibitem{kramer2019topological}
R.~Kramer, A.~Popolitov, and S.~Shadrin, ``Topological recursion for monotone
  orbifold {Hurwitz} numbers: a proof of the {Do}-{Karev} conjecture,'' {\em
  arXiv preprint arXiv:1909.02302}, 2019.

\bibitem{dunin2019loop}
P.~Dunin-Barkowski, R.~Kramer, A.~Popolitov, and S.~Shadrin, ``Loop equations
  and a proof of {Zvonkine}'s $ qr $-{E}{L}{S}{V} formula,'' {\em arXiv
  preprint arXiv:1905.04524}, 2019.

\bibitem{dunin2019combinatorial}
P.~Dunin-Barkowski, A.~Popolitov, S.~Shadrin, and A.~Sleptsov, ``Combinatorial
  structure of colored {H}{O}{M}{F}{L}{Y}-{P}{T} polynomials for torus knots,''
  {\em Communications in Number Theory and Physics}, vol.~13, no.~4,
  pp.~763--826, 2019.

\bibitem{borot2017blobbed}
G.~Borot and S.~Shadrin, ``Blobbed topological recursion: properties and
  applications,'' in {\em Mathematical Proceedings of the Cambridge
  Philosophical Society}, vol.~162, pp.~39--87, Cambridge University Press,
  2017.

\bibitem{bychkov2020combinatorics}
B.~Bychkov, P.~Dunin-Barkowski, and S.~Shadrin, ``Combinatorics of
  {Bousquet}-{M}{\'e}lou--{Schaeffer} numbers in the light of topological
  recursion,'' {\em European Journal of Combinatorics}, vol.~90, p.~103184,
  2020.

\bibitem{its1990differential}
A.~Its, A.~Izergin, V.~Korepin, and N.~Slavnov, ``Differential equations for
  quantum correlation functions,'' {\em International Journal of Modern Physics
  B}, vol.~4, no.~05, pp.~1003--1037, 1990.

\bibitem{izergin1992determinant}
A.~G. Izergin, D.~A. Coker, and V.~E. Korepin, ``Determinant formula for the
  six-vertex model,'' {\em Journal of Physics A: Mathematical and General},
  vol.~25, no.~16, p.~4315, 1992.

\bibitem{gorsky1995integrability}
A.~Gorsky, I.~Krichever, A.~Marshakov, A.~Mironov, and A.~Morozov,
  ``Integrability and {Seiberg}-{Witten} exact solution,'' {\em Physics Letters
  B}, vol.~3, no.~355, pp.~466--474, 1995.

\bibitem{gorsky1998multiscale}
A.~Gorsky, S.~Gukov, and A.~Mironov, ``Multiscale {N}= 2 {S}{U}{S}{Y} field
  theories, integrable systems and their stringy/brane origin,'' {\em Nuclear
  Physics B}, vol.~517, no.~1-3, pp.~409--461, 1998.

\bibitem{date1982transformation}
E.~Date, M.~Jimbo, M.~Kashiwara, and T.~Miwa, ``Transformation groups for
  soliton equations,'' {\em Publications of the Research Institute for
  Mathematical Sciences}, vol.~18, no.~3, pp.~1077--1110, 1982.

\bibitem{jimbo1983solitons}
M.~Jimbo and T.~Miwa, ``Solitons and infinite dimensional {Lie} algebras,''
  {\em Publications of the Research Institute for Mathematical Sciences},
  vol.~19, no.~3, pp.~943--1001, 1983.

\bibitem{sato1983soliton}
M.~Sato, ``Soliton equations as dynamical systems on infinite dimensional
  {Grassmann} manifold,'' in {\em North-Holland Mathematics Studies}, vol.~81,
  pp.~259--271, Elsevier, 1983.

\bibitem{segal1985loop}
G.~Segal and G.~Wilson, ``Loop groups and equations of {K}d{V} type,'' {\em
  Publications Math{\'e}matiques de l'Institut des Hautes {\'E}tudes
  Scientifiques}, vol.~61, no.~1, pp.~5--65, 1985.

\bibitem{brezin1993exactly}
E.~Brezin and V.~Kazakov, ``Exactly solvable field theories of closed
  strings,'' in {\em The Large N Expansion In Quantum Field Theory And
  Statistical Physics: From Spin Systems to 2-Dimensional Gravity},
  pp.~711--717, World Scientific, 1993.

\bibitem{douglas1990strings}
M.~R. Douglas and S.~H. Shenker, ``Strings in less than one dimension,'' {\em
  Nuclear Physics B}, vol.~335, no.~3, pp.~635--654, 1990.

\bibitem{gross1993nonperturbative}
D.~J. Gross and A.~A. Migdal, ``A nonperturbative treatment of two-dimensional
  quantum gravity,'' in {\em The Large N Expansion In Quantum Field Theory And
  Statistical Physics: From Spin Systems to 2-Dimensional Gravity},
  pp.~742--774, World Scientific, 1993.

\bibitem{KHARCHEV_1995}
S.~Kharchev, A.~Marshakov, A.~Mironov, and A.~Morozov, ``Generalized
  {Kazakov}-{Migdal}-{Kontsevich} {Model}: group theory aspects,'' {\em
  International Journal of Modern Physics A}, vol.~10, p.~2015–2051, Jun
  1995.

\bibitem{mironov1996unitary}
A.~Mironov, A.~Morozov, and G.~W. Semenoff, ``Unitary matrix integrals in the
  framework of the generalized {Kontsevich} model,'' {\em International Journal
  of Modern Physics A}, vol.~11, no.~28, pp.~5031--5080, 1996.

\bibitem{mironov2013genus}
A.~Mironov, A.~Morozov, and A.~Sleptsov, ``On genus expansion of knot
  polynomials and hidden structure of {Hurwitz} tau-functions,'' {\em The
  European Physical Journal C}, vol.~73, no.~7, p.~2492, 2013.

\bibitem{Okounkov_2000}
A.~Okounkov, ``Toda equations for {Hurwitz} numbers,'' {\em Mathematical
  Research Letters}, vol.~7, no.~4, p.~447–453, 2000.

\bibitem{takasaki1992quasi}
K.~Takasaki and T.~Takebe, ``Quasi-classical limit of {K}{P} hierarchy,
  {W}-symmetries and free fermions,'' {\em arXiv preprint hep-th/9207081},
  1992.

\bibitem{krichever1992dispersionless}
I.~Krichever, ``The dispersionless {Lax} equations and topological minimal
  models,'' {\em Communications in mathematical physics}, vol.~143, no.~2,
  pp.~415--429, 1992.

\bibitem{dubrovin1992hamiltonian}
B.~Dubrovin, ``Hamiltonian formalism of {Whitham}-type hierarchies and
  topological {Landau}-{Ginsburg} models,'' {\em Communications in mathematical
  physics}, vol.~145, no.~1, pp.~195--207, 1992.

\bibitem{kontsevich1992intersection}
M.~Kontsevich, ``Intersection theory on the moduli space of curves and the
  matrix {Airy} function,'' {\em Communications in Mathematical Physics},
  vol.~147, no.~1, pp.~1--23, 1992.

\bibitem{gross1980possible}
D.~J. Gross and E.~Witten, ``Possible third-order phase transition in the
  large-{N} lattice gauge theory,'' {\em Physical Review D}, vol.~21, no.~2,
  p.~446, 1980.

\bibitem{brezin1980external}
E.~Br{\'e}zin and D.~J. Gross, ``The external field problem in the large {N}
  limit of {Q}{C}{D},'' {\em Physics Letters B}, vol.~97, no.~1, pp.~120--124,
  1980.

\bibitem{alexandrov2012integrability}
A.~Alexandrov, A.~Mironov, A.~Morozov, and S.~Natanzon, ``Integrability of
  {Hurwitz} partition functions,'' {\em Journal of Physics A: Mathematical and
  Theoretical}, vol.~45, no.~4, p.~045209, 2012.

\bibitem{gerasimov1991matrix}
A.~Gerasimov, A.~Marshakov, A.~Mironov, A.~Morozov, and A.~Orlov, ``Matrix
  models of two-dimensional gravity and {Toda} theory,'' {\em Nuclear Physics
  B}, vol.~357, no.~2-3, pp.~565--618, 1991.

\bibitem{kharchev1991matrix}
S.~Kharchev, A.~Marshakov, A.~Mironov, A.~Orlov, and A.~Zabrodin, ``Matrix
  models among integrable theories: {Forced} hierarchies and operator
  formalism,'' {\em Nuclear Physics B}, vol.~366, no.~3, pp.~569--601, 1991.

\bibitem{kharchev1993generalized}
S.~Kharchev, A.~Marshakov, A.~Mironov, and A.~Morozov, ``Generalized
  {Kontsevich} model versus {Toda} hierarchy and discrete matrix models,'' {\em
  Nuclear Physics B}, vol.~397, no.~1-2, pp.~339--378, 1993.

\bibitem{witten1990two}
E.~Witten, ``Two-dimensional gravity and intersection theory on moduli space,''
  {\em Surveys in differential geometry}, vol.~1, no.~1, pp.~243--310, 1990.

\bibitem{Kharchev_1992}
S.~Kharchev, A.~Marshakov, A.~Mironov, A.~Morozov, and A.~Zabrodin,
  ``Unification of all string models with c<1,'' {\em Physics Letters B},
  vol.~275, p.~311–314, Jan 1992.

\bibitem{Alexandrov_2009}
A.~Alexandrov, A.~Mironov, and A.~Morozov, ``{B}{G}{W}{M} as second constituent
  of complex matrix model,'' {\em Journal of High Energy Physics}, vol.~2009,
  p.~053–053, Dec 2009.

\bibitem{rusakov1990loop}
B.~Y. Rusakov, ``Loop averages and partition functions in {U}({N}) gauge theory
  on two-dimensional manifolds,'' {\em Modern Physics Letters A}, vol.~5,
  no.~09, pp.~693--703, 1990.

\bibitem{gross1993two}
D.~J. Gross and W.~Taylor~IV, ``Two-dimensional {Q}{C}{D} is a string theory,''
  {\em Nuclear Physics B}, vol.~400, no.~1-3, pp.~181--208, 1993.

\bibitem{kostov1997two}
I.~K. Kostov and M.~Staudacher, ``Two-dimensional chiral matrix models and
  string theories,'' {\em Physics Letters B}, vol.~394, no.~1-2, pp.~75--81,
  1997.

\bibitem{kimura2008holomorphic}
Y.~Kimura and S.~Ramgoolam, ``Holomorphic maps and the complete 1/{N} expansion
  of 2{D} {S}{U}({N}) {Yang}-{Mills},'' {\em Journal of High Energy Physics},
  vol.~2008, no.~06, p.~015, 2008.

\bibitem{griguolo2005double}
L.~Griguolo, D.~Seminara, and R.~Szabo, ``Double scaling string theory of
  {Q}{C}{D} in two dimensions,'' {\em Fortschritte der Physik: Progress of
  Physics}, vol.~53, no.~5-6, pp.~615--620, 2005.

\bibitem{douglas1993large}
M.~R. Douglas and V.~A. Kazakov, ``Large {N} phase transition in continuum
  {Q}{C}{D}$_{2}$,'' {\em Physics letters B}, vol.~319, no.~1-3, pp.~219--230,
  1993.

\bibitem{mironov2013genus2}
A.~D. Mironov, A.~Y. Morozov, and A.~V. Sleptsov, ``Genus expansion of
  {H}{O}{M}{F}{L}{Y} polynomials,'' {\em Theoretical and Mathematical Physics},
  vol.~177, no.~2, pp.~1435--1470, 2013.

\bibitem{sleptsov2014hidden}
A.~Sleptsov, ``Hidden structures of knot invariants,'' {\em International
  Journal of Modern Physics A}, vol.~29, no.~29, p.~1430063, 2014.

\bibitem{mironov2013character}
A.~Mironov, A.~Morozov, and A.~Morozov, ``Character expansion for
  {H}{O}{M}{F}{L}{Y} polynomials {I}: {Integrability} and difference
  equations,'' in {\em Strings, gauge fields, and the geometry behind: the
  legacy of Maximilian Kreuzer}, pp.~101--118, World Scientific, 2013.

\bibitem{okounkov2006gromov}
A.~Okounkov and R.~Pandharipande, ``Gromov-{Witten} theory, {Hurwitz} theory,
  and completed cycles,'' {\em Annals of mathematics}, pp.~517--560, 2006.

\bibitem{Mironov_2017}
A.~Mironov and A.~Morozov, ``On the complete perturbative solution of
  one-matrix models,'' {\em Physics Letters B}, vol.~771, p.~503–507, Aug
  2017.

\bibitem{macdonald1998symmetric}
I.~G. Macdonald, {\em Symmetric functions and {Hall} polynomials}.
\newblock Oxford university press, 1998.

\bibitem{kac2013bombay}
V.~G. Kac, A.~K. Raina, and N.~Rozhkovskaya, {\em Bombay lectures on highest
  weight representations of infinite dimensional {Lie} algebras}, vol.~29.
\newblock World scientific, 2013.

\bibitem{miwa2000solitons}
T.~Miwa, M.~Jimbo, and E.~Date, {\em Solitons: {Differential} equations,
  symmetries and infinite dimensional algebras}, vol.~135.
\newblock Cambridge University Press, 2000.

\bibitem{orlov2001hypergeometric}
A.~Y. Orlov and D.~Scherbin, ``Hypergeometric solutions of soliton equations,''
  {\em Theoretical and Mathematical Physics}, vol.~128, no.~1, pp.~906--926,
  2001.

\bibitem{bouchard2007hurwitz}
V.~Bouchard and M.~Marino, ``Hurwitz numbers, matrix models and enumerative
  geometry,'' in {\em Proc. Symp. Pure Math.}, vol.~78, pp.~263--283, 2007.

\bibitem{mironov2009virasoro}
A.~Mironov and A.~Morozov, ``Virasoro constraints for {Kontsevich}-{Hurwitz}
  partition function,'' {\em Journal of High Energy Physics}, vol.~2009,
  no.~02, p.~024, 2009.

\bibitem{morozov2018q}
A.~Morozov, A.~Popolitov, and S.~Shakirov, ``On (q, t)-deformation of
  {Gaussian} matrix model,'' {\em Physics Letters B}, vol.~784, pp.~342--344,
  2018.

\bibitem{mironov2017complete}
A.~Mironov and A.~Morozov, ``On the complete perturbative solution of
  one-matrix models,'' {\em Physics Letters B}, vol.~771, pp.~503--507, 2017.

\bibitem{mironov2017correlators}
A.~Mironov and A.~Morozov, ``Correlators in tensor models from character
  calculus,'' {\em Physics Letters B}, vol.~774, pp.~210--216, 2017.

\bibitem{mironov2018sum}
A.~Mironov and A.~Morozov, ``Sum rules for characters from
  character-preservation property of matrix models,'' {\em Journal of High
  Energy Physics}, vol.~2018, no.~8, p.~163, 2018.

\bibitem{cordova2018orbifolds}
C.~C{\'o}rdova, B.~Heidenreich, A.~Popolitov, and S.~Shakirov, ``Orbifolds and
  exact solutions of strongly-coupled matrix models,'' {\em Communications in
  Mathematical Physics}, vol.~361, no.~3, pp.~1235--1274, 2018.

\bibitem{fulton2013representation}
W.~Fulton and J.~Harris, {\em Representation theory: a first course}, vol.~129.
\newblock Springer Science \& Business Media, 2013.

\bibitem{Alexandrov_2013}
A.~Alexandrov, ``From {Hurwitz} numbers to {Kontsevich}–{Witten}
  tau-function: {A} connection by {Virasoro} operators,'' {\em Letters in
  Mathematical Physics}, vol.~104, p.~75–87, Aug 2013.

\bibitem{morozov2010unitary}
A.~Y. Morozov, ``Unitary integrals and related matrix models,'' {\em
  Theoretical and Mathematical Physics}, vol.~162, no.~1, pp.~1--33, 2010.

\bibitem{Alexandrov_2018}
A.~Alexandrov, ``Cut-and-join description of generalized
  {Brezin}–{Gross}–{Witten} model,'' {\em Advances in Theoretical and
  Mathematical Physics}, vol.~22, no.~6, p.~1347–1399, 2018.

\end{thebibliography}
	
\end{document}